\documentclass[slac_one]{revtex4}
\usepackage{graphicx}
\usepackage{fancyhdr}
\newcommand{\zp}{\ensuremath{Z^{\prime}}}
\newcommand{\pt}{\ensuremath{p_{T}}}
\newcommand{\et}{\ensuremath{E_{T}}}
\newcommand{\met}{\ensuremath{E_{T}\!\!\!\!\!\!/}}
\newcommand{\gev}{\ensuremath{\rm GeV}}
\newcommand{\gevc}{\ensuremath{{\rm GeV}/c}}
\newcommand{\gevcsq}{\ensuremath{{\rm GeV}/c^2}}

\newcommand{\fbarn} {\ensuremath{\mathrm{fb^{-1}}}}
\newcommand{\g} {\ensuremath{\gamma}}
\def \kaione {{\tilde \chi}_1^0}
\def \kaioneplus {{\tilde \chi}_1^+}
\def \kaitwo {{\tilde \chi}_2^0}
\def\stop{\tilde{t}}
\begin{document}
\title{{\small{Hadron Collider Physics Symposium (HCP2008),
Galena, Illinois, USA}}\\ 
\vspace{12pt}
Other Beyond Standard Model Searches at the Tevatron} 

%

\author{Shin-Shan Yu (on behalf of the CDF and D0 Collaborations)}
\affiliation{Fermi National Accelerator Laboratory, P.O.~Box~500, Batavia, IL~60510, USA}
%


\begin{abstract}
 We present the results of searches for non-standard model phenomena, with 
focus on signature-based searches and searches driven by non-supersymmetry 
(non-SUSY) models. The analyses use 1.0--2.5~\fbarn\ of data from $p\bar{p}$ 
collisions at $\sqrt{s}=1.96$ TeV collected with the CDF and D0 detectors at 
the Fermilab Tevatron. No significant excess in data has been observed. We 
report on the event counts, kinematic distributions, and limits on selected 
model parameters. 

\end{abstract}

\maketitle

\thispagestyle{fancy}


\section{INTRODUCTION} 
The standard model (SM) of elementary particle physics describes the 
structure of fundamental particles and how they interact via gauge bosons.
To date, almost all experimental results have agreed with the prediction by 
the standard model. However, many questions can be raised, which indicate 
that the standard model is not complete. For example, ``Why is there a 
hierarchy between the electroweak scale (1~TeV) and the gravitational scale 
($10^{16}$~TeV)?'', ``What are the origins of mass?'', ``Why is there a 
spectrum of fermion masses? Are there only three generations?'', {\it etc}. 
Although the most popular extension of the standard model is supersymmetry 
(SUSY), there are other equally well-motivated models, such as extra dimension,
 compositeness, 4$^\mathrm{th}$ generation, technicolor, {\it etc}. In this 
document, we present results of signature-based searches and searches inspired 
by non-SUSY models, using 1.0--2.5~\fbarn\ of data collected with the CDF and 
D0 detectors. In signature-based searches, we apply generic selection 
criteria in order to be sensitive to a wide range of new physics. We report on 
the event counts and various kinematic distributions of data and predicted 
backgrounds. In model-inspired searches, we optimize selection criteria to 
obtain the best sensitivity for selected models. If no significant excess is 
found, we report limits on model parameters.

\section{RESULTS OF BEYOND STANDARD MODEL SEARCHES AT THE TEVATRON}

\subsection{\boldmath Search for Anomalous Production of $\g bj\met$ 
            \label{sec:gbjm}}
\begin{figure}
\begin{tabular}{cc}
\includegraphics[width=0.4\textwidth]{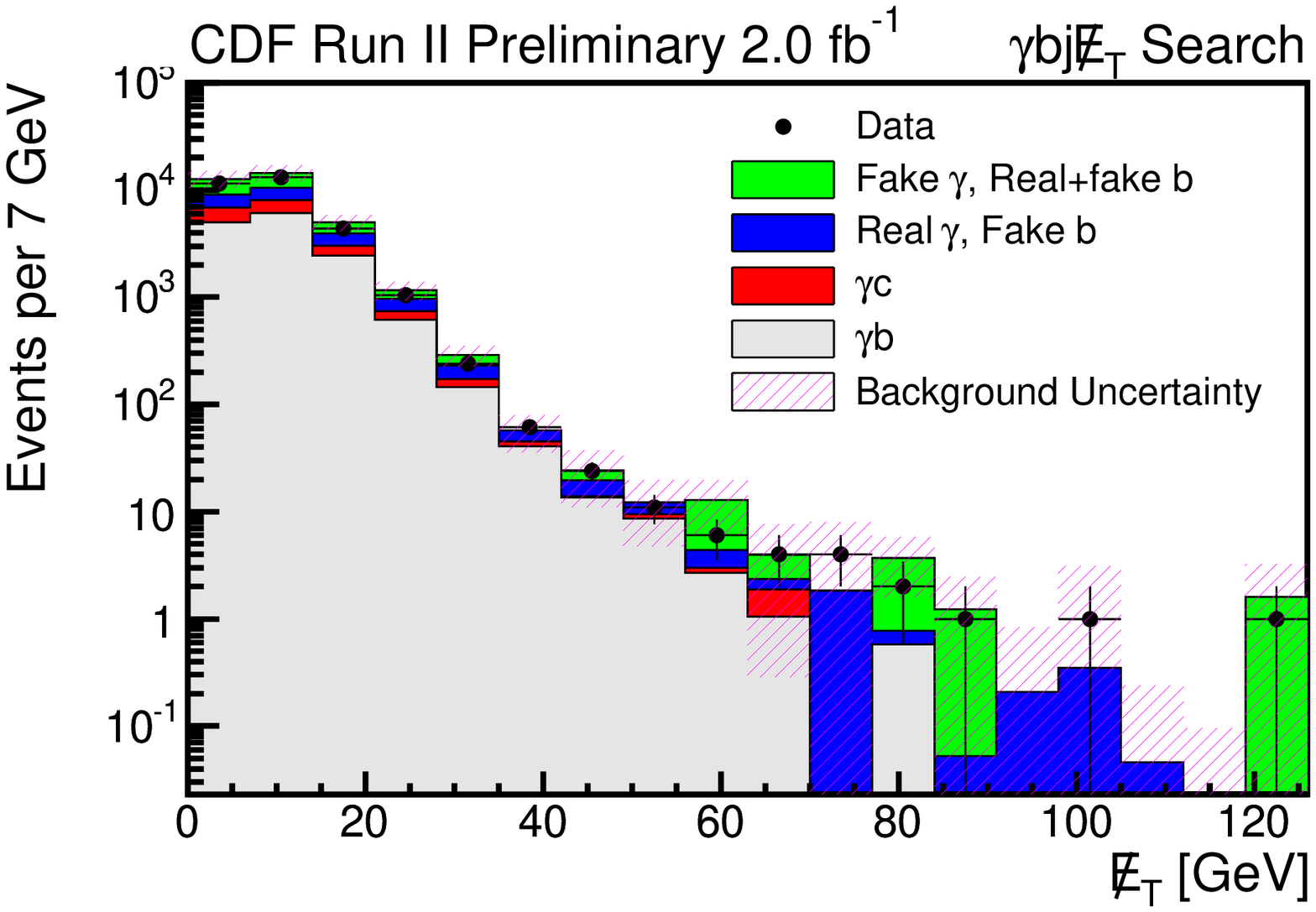} &   
\includegraphics[width=0.4\textwidth]{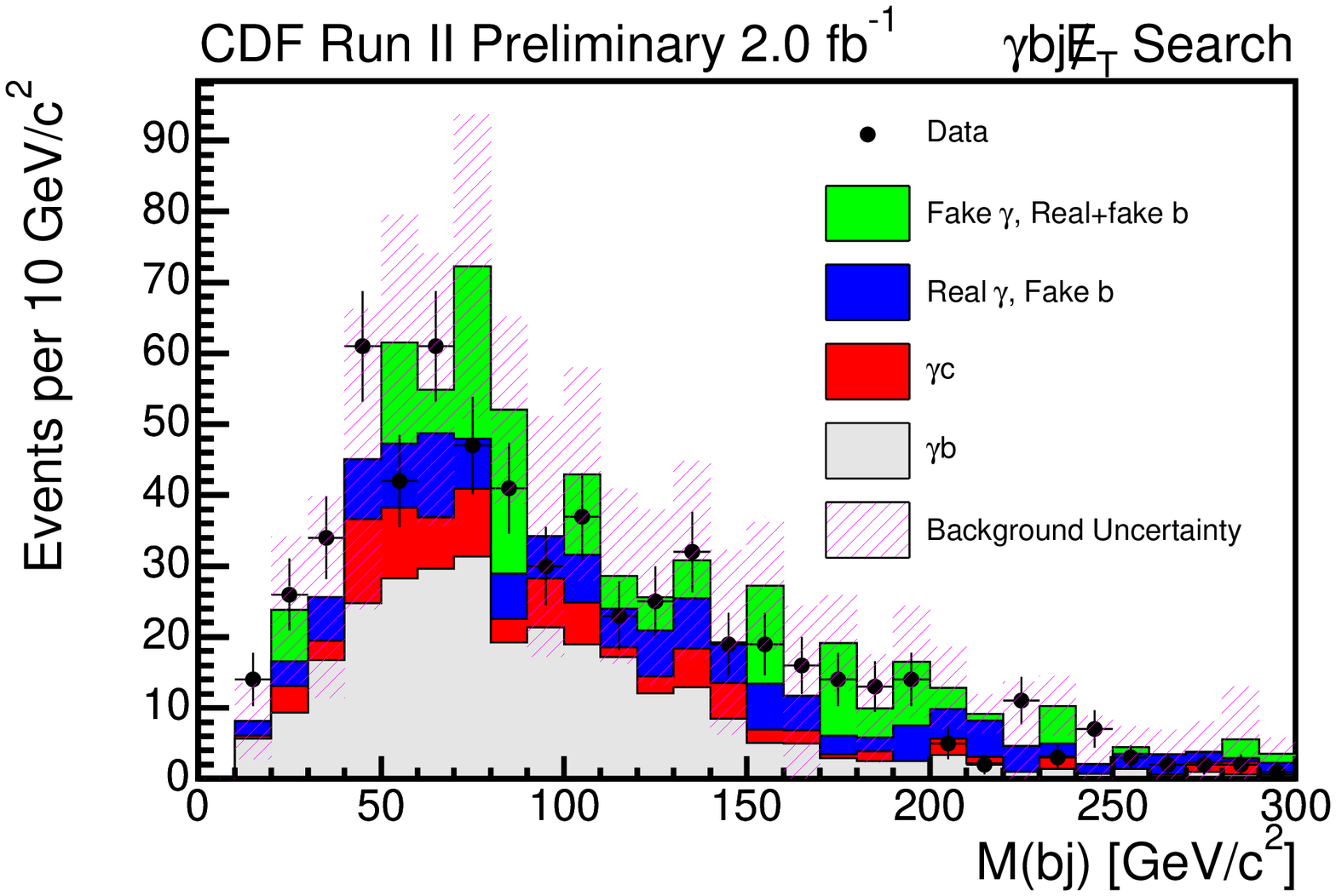} \\
\end{tabular}  
	\caption{\label{fig:gbjmet}
	CDF search for anomalous production of $\g bj\met\;$:
	the $\met\;$ (left) and $M_{bj}$ (right) distributions observed 
	(markers) and background prediction (filled histograms). 
	The hatched-region indicates the total 
	uncertainty on the predicted background in each bin.	
	}
\end{figure}

\unboldmath
The CDF collaboration has performed a signature-based search in the inclusive 
$\g bj\met\;$ final state using 2.0~\fbarn\ of data. The $\g bj\met\;$ 
signature raised great interest for two main reasons. First, this final state 
has been predicted by several SUSY models\footnote{These models had 
been proposed to explain the CDF $ee\g\g\met\;$ event observed in Run I~\cite{Toback_all}.}\cite{Kane:1996ny,Kane}, {\it e.g.}, the production of a chargino and a 
neutralino, when $\kaitwo$ is photino-like and the LSP $\kaione$ is 
Higgsino-like, via the decay chain: 
$\kaioneplus \kaitwo \rightarrow  (\bar{b}\stop )(\gamma\kaione) \rightarrow 
(\bar{b}c \kaione)(\gamma\kaione)  \rightarrow (\gamma \bar{b} c \met\;)
$. Second, the dominant backgrounds are mis-identifications of either the 
photon or the $b$-quark candidates and mismeasurements of the jet energy which 
induce $\met\;$ not associated with unobserved neutral particles (fake $\met\;$).  
The SM processes which produce real $\g bj\met\;$ are expected to contribute at
 most 2\%. Therefore, a significant excess in data will be an indication of new
 physics. Events are required to have 
a central\footnote{Throughout this document, all central objects have 
detector pseudo-rapidity $\left|\eta^\mathrm{det}\right| < 1.1$.} photon with 
transverse energy $\et>25~\gev$, at least two jets with $\et>15~\gev$ and 
$\left|\eta^\mathrm{det}\right| < 2.0$, at least one of the jets must be 
identified as originating from a $b$ quark (``$b$-tagged'') using the tight 
SECVTX algorithm~\cite{SECVTX}, and missing transverse energy 
$\met\; > 25~\gev$. Figure~\ref{fig:gbjmet} shows the $\met\;$ and dijet mass 
$M_{bj}$ distributions from data and predicted background. 
Other kinematic distributions, such as jet multiplicity, \et\ of photon, 
\et\ of $b$-tagged jet, {\it etc.}, have also been examined and no significant 
excess has been found. 
The observed number of events in data is 617, which is consistent 
with the expected number of background events, $637\pm 139$. 

\subsection{\boldmath Search for Anomalous Production of $\ell\g b\met\;\;$ 
  and Measurement of SM $t\bar{t}\g$ Production Cross-section
  \label{sec:ttg}} 
\begin{figure}
\includegraphics[width=0.28\textwidth, angle=90]{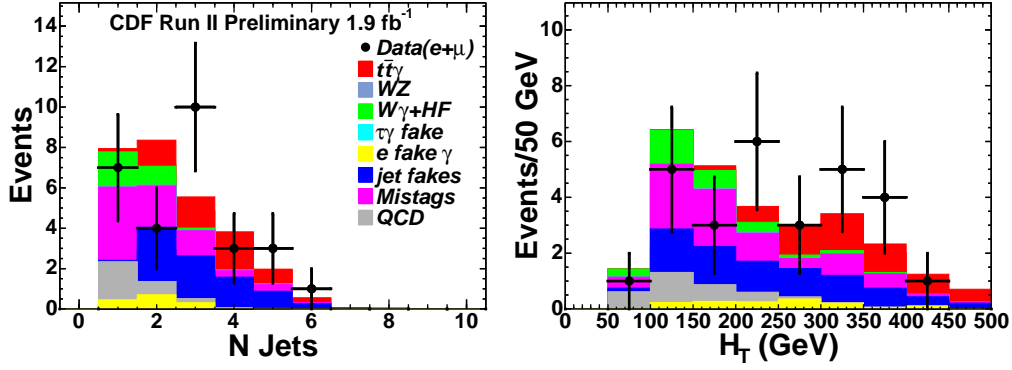}   
	\caption{\label{fig:ttg}
	CDF search for anomalous production of $\ell\g b\met\;$: 
	the jet multiplicity (left) and $H_T$ (right) distributions observed 
	(markers) and background prediction (filled histograms).  
	The contribution of SM $t\bar{t}\g$ increases as the jet multiplicity 
	and $H_T$ increase. }
\end{figure}

\unboldmath
Ref.~\cite{Kane:1996ny} predicts in the Minimal Supersymmetric 
Standard Model (MSSM) an exotic decay channel of the top quark, which may 
compete with $t\rightarrow Wb$, into a light stop and a light Higgsino-like 
neutralino. 
A $t\bar{t}$ pair may then decay via $t\bar{t}\rightarrow Wb\tilde t \tilde
 \chi_i \rightarrow \ell\bar{\nu}_{\ell}bc \kaione \kaione \g + X$. Instead of 
searching for this MSSM model only, the CDF collaboration has performed a 
model-independent search in the inclusive $\ell\g b\met\;$ final state using 
1.9~\fbarn\ of data, where $\ell$ is an electron or a muon. 
Since this signature is rare, the \et\ and $b$-tagging requirements are 
looser than those in Section~\ref{sec:gbjm}: a central electron or muon with 
$\pt>20~\gev$, a central photon with $\et>10~\gev$, at least one jet which is 
$b$-tagged by the loose SECVTX algorithm~\cite{SECVTX}, and $\met\; > 20~\gev$.
 Figure~\ref{fig:ttg} shows the jet multiplicity and $H_T$\footnote{The 
$H_T$ is defined as the scalar sum \pt\ of all identified objects in an event.}
 distributions from the inclusive $\ell\g b\met\;$ final state. No significant 
excess in data is found: 28 observed and $27.9 {+3.6 \atop -3.5}$ 
expected. The background has a significant 
contribution from the SM $t\bar{t}\g$ production, especially in the lepton + 
jets channel. After requiring $H_T>200~\gev$ and two additional jets 
($\geq 3$ jets with $\geq$ 1 $b$-tag in total), the $t\bar{t}\g$ cross-section 
has been measured to be $0.15\pm0.08~\rm{pb}$, which is consistent with the 
next-to-leading-order (NLO) prediction, $0.080\pm0.012~\rm{pb}$~\cite{sigma_ttg}.

\subsection{\boldmath Search for Anomalous Production of $\g\g\met\;$ \label{sec:ggmet}} 
\begin{figure}
\begin{tabular}{cc}
\includegraphics[width=0.4\textwidth]{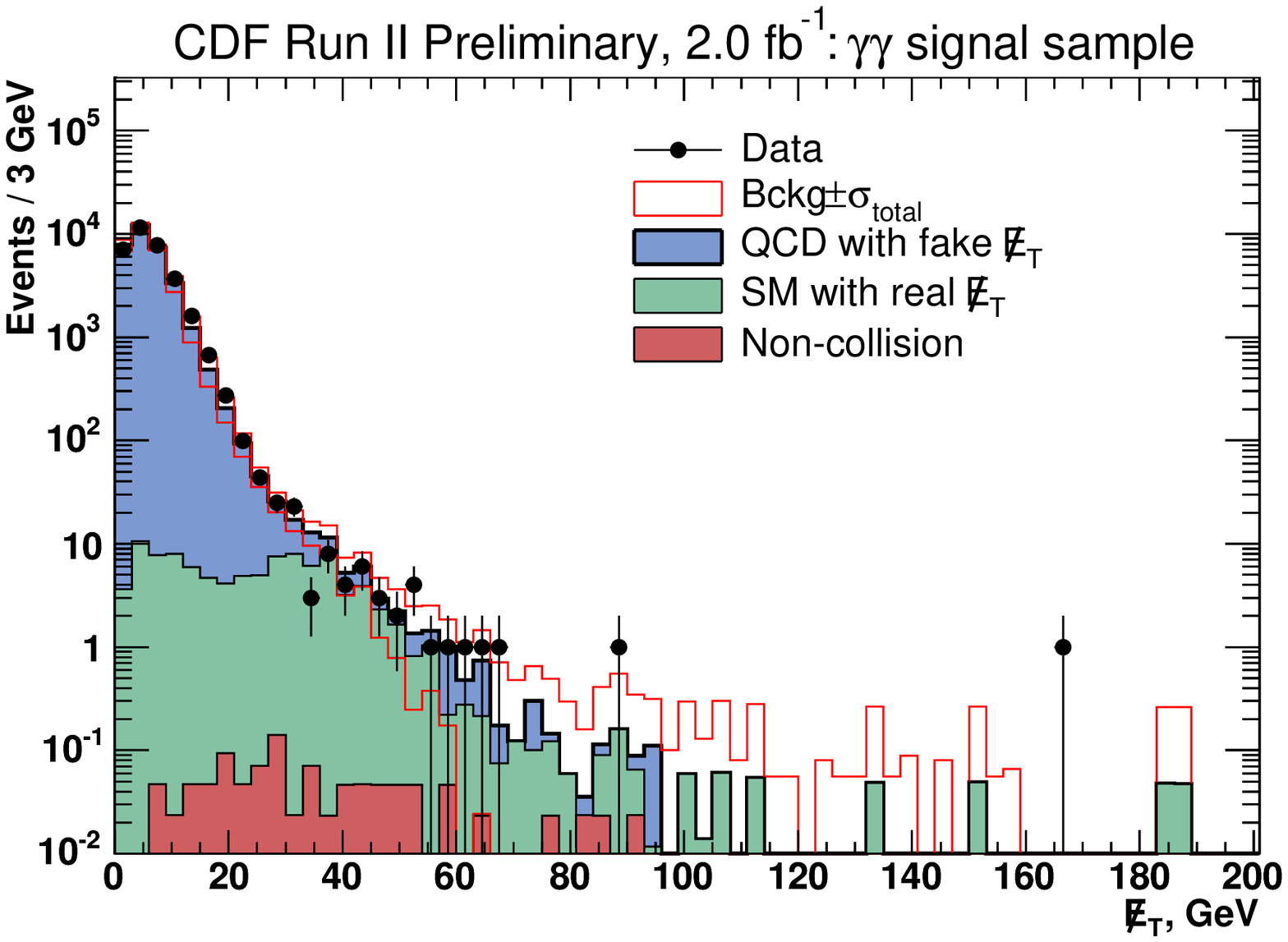} &   
\includegraphics[width=0.4\textwidth]{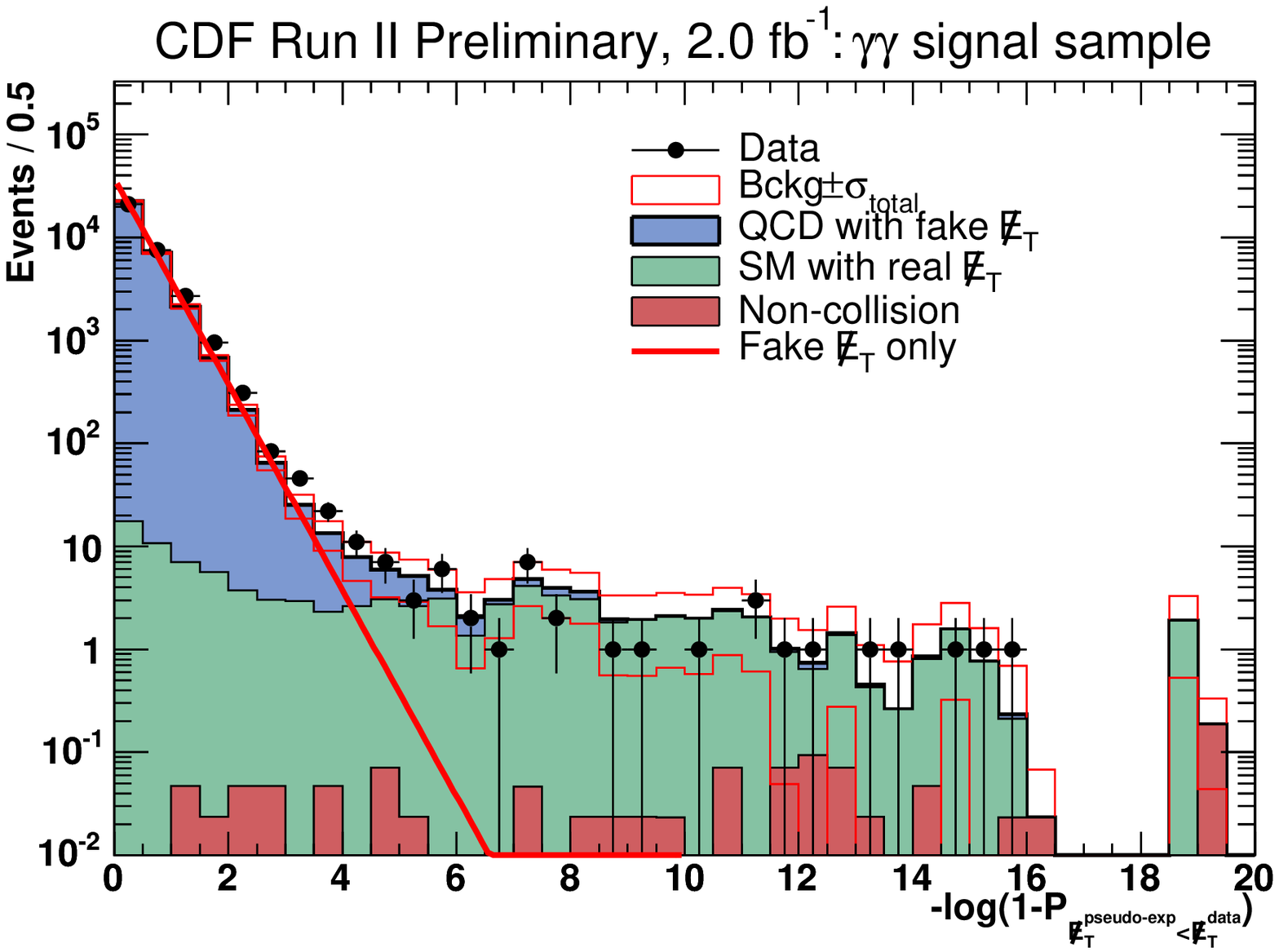} \\\end{tabular}  
	\caption{\label{fig:ggmet}
	CDF search for anomalous production of $\g\g\met\;$:
	the $\met\;$ (left) and $\met\;$ significance (right) distributions 
	observed (markers) and background prediction (filled histograms). The 
	$\met\;$ significance is defined as 
	$-\log(1-{\cal P}_{\met\;^\mathrm{pseudo-exp}<\met\;^{data}})$, namely
 	how often the observed $\met\;$ is larger than a $\met\;$ value which 
	is randomly picked from the predicted fake $\met\;$ distribution. }
\end{figure}

\unboldmath
Anomalous production of inclusive $\g\g\met\;$ events has been predicted by 
many models, such as gauge-mediated SUSY breaking~\cite{ggMET}, fermiophobic 
Higgs~\cite{fpH}, 4$^\mathrm{th}$ generation~\cite{fourth}, and the $E_6$ 
model~\cite{E6ggmet}. The CDF collaboration has carried out a signature-based 
search using 2.0~\fbarn\ of data. Two central photons with $\et > 13~\gev$ 
are required. The non-collision backgrounds from beam halos and cosmic rays 
are suppressed by requiring photons to be in time with a $p\bar{p}$ 
collision, where the photon time is measured with a novel timing system 
(EM Timing)~\cite{EMTiming}. 
Instead of making a tight requirement on $\met\;$, this analysis selects events
 with large ``$\met\;$ significance''. A data-based model predicts the fake 
$\met\;$ distribution induced by mis-measurement of jet energies and soft 
unclustered energies\footnote{The soft unclustered energies refer to energies 
not included by jet reconstruction algorithms and are from underlying events 
or multiple interactions.} and calculates the $\met\;$ significance event by 
event. Figure~\ref{fig:ggmet} shows the distributions of $\met\;$ and $\met\;$ 
significance in the diphoton sample. A minimum requirement on the $\met\;$ 
significance removes events with large, fake $\met\;$ and keeps a good 
acceptance for events with small, real $\met\;$ which would have been rejected 
by a straight $\met\;$ cut. For $\met\;$ significance greater than 5, 34 events
 are observed in data, which is consistent with the background expectation, 
$48.6\pm7.5$. Note that the QCD multi-jet or diphoton+jet events are largely 
removed and the events selected are mostly SM $W\gamma$ events with real 
$\met\;$\footnote{Here, the lepton from $W$ is misidentified as one of the 
photons.}.

\subsection{Model-independent Global Search for New Physics}
\begin{figure}
\begin{tabular}{cc}
\includegraphics[width=0.3\textwidth, angle=270]{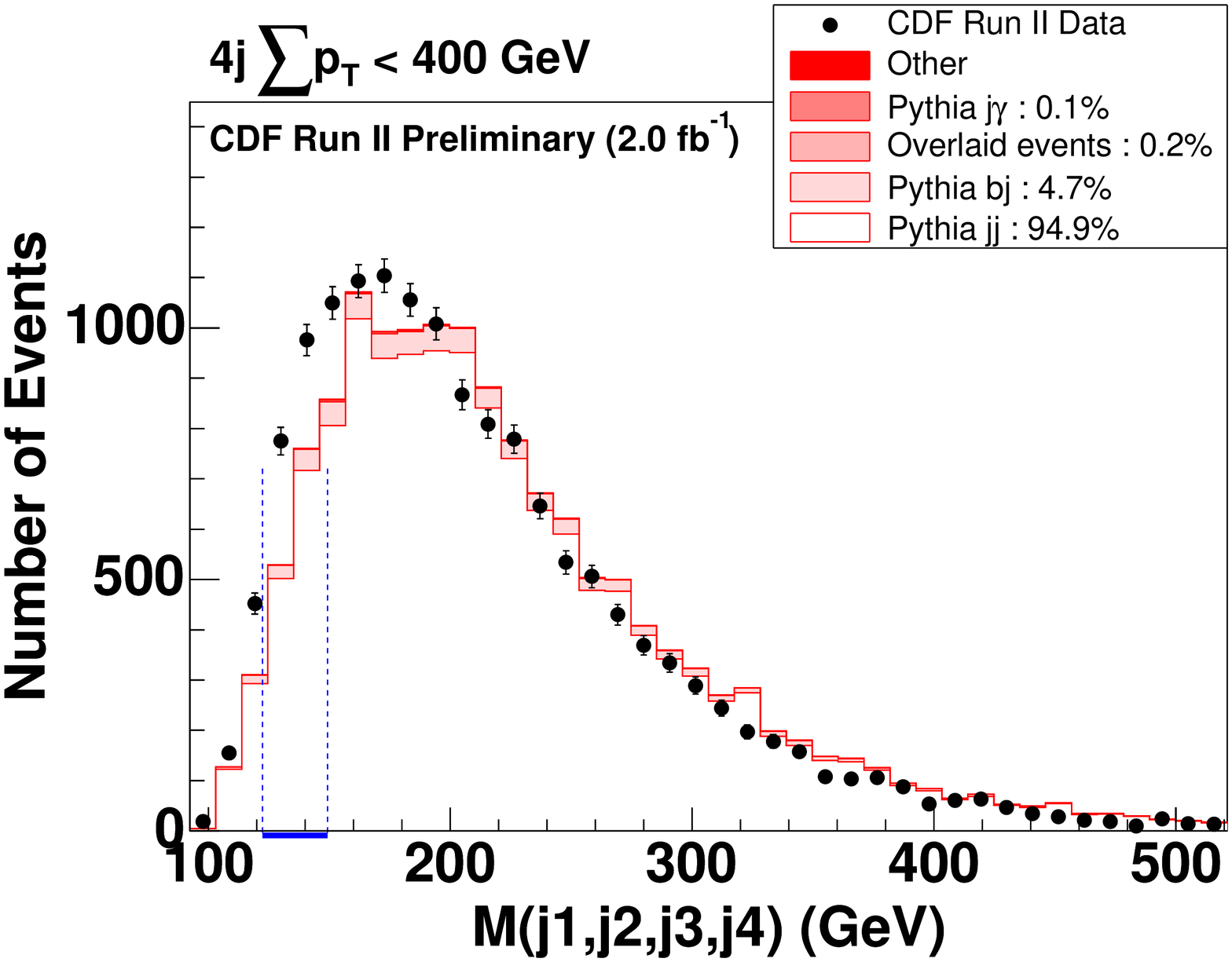} &   
\includegraphics[width=0.3\textwidth, angle=270]{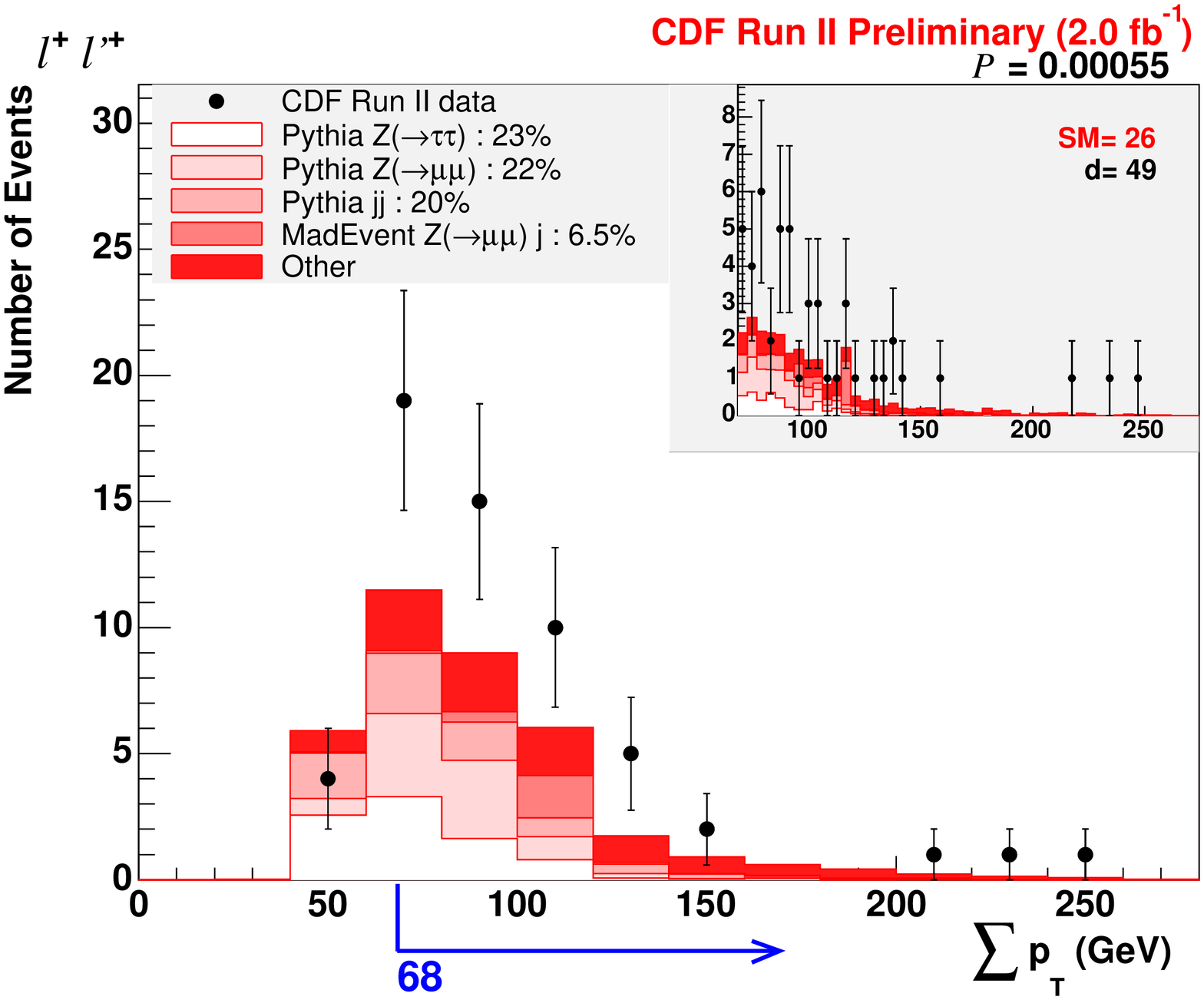} \\
\end{tabular} 
	\caption{\label{fig:global}
	CDF model-independent global search. The left figure shows the only 
	significant bump found by {\sc Bump Hunter}, the invariant mass of 
	all four jets in the 4-jet final state, as indicated by the blue 
	dashed lines. The right figure shows the final state with the most 
	significant excess in the $\Sigma \pt$ distribution found by 
	{\sc SLEUTH}, same-sign dilepton with different flavors and 
	$\Sigma \pt > 68~\gevc$.}
\end{figure}

The CDF collaboration has performed a model-independent global search in 
2.0~\fbarn\ of data which contain over four 
million high-\pt\ events~\cite{global,Aaltonen:2007dg}. This global search has 
three algorithms: {\sc VISTA}, {\sc  Bump Hunter}, and {\sc SLEUTH}, and aims 
to look for new physics in every possible final state without bias toward 
any new physics model.  
The first algorithm, {\sc VISTA}, searches for discrepancies in the 
total event counts and shapes of kinematic distributions. 
Data are partitioned into 399 exclusive final states according to 
combinations of detectable objects: \g, $e$, $\mu$, $\tau$, $b$-jet, jet, 
and $\met\;$. All objects are required to have $\pt \geq 17~\gevc$. The 
background prediction is estimated with Monte Carlo (MC) using standard HEP 
event generators and CDF detector simulation. The $k$-factors for the SM 
cross-sections and the data to MC scale factors for the object efficiencies 
and mis-identification probabilities are determined from data by a global fit 
to all final states. After accounting for the trials factor associated with 
looking at so many final states, no significant discrepancy is found in the 
event counts, but 555 out of 19,650 kinematic distributions have significant 
different shapes between data and background prediction. Careful investigations
 show that these discrepancies are attributed to the difficulty in modeling 
soft QCD jet radiation in the simulation. The second algorithm, 
{\sc Bump Hunter}, searches for narrow resonances in invariant mass 
distributions. The search window is defined based on the expected detector 
resolution. Out of 5036 invariant mass distributions, the only significant 
bump found is the invariant mass of all four jets in the 4-jet final state 
(see Figure~\ref{fig:global}). 
However, this bump arises from the same, imperfect modeling of soft QCD jets 
seen in {\sc VISTA}. The third algorithm, {\sc SLEUTH}, assumes new physics 
appears as excess in the tail of scalar sum \pt\ ($\Sigma \pt$) 
distributions. For each final state, {\sc SLEUTH} determines the 
semi-infinite region of $\Sigma \pt$ which has the most significant excess in 
data. Figure~\ref{fig:global} shows the final state with the most significant 
region. After taking into account the trials factor, $\sim$ 8\% of 
hypothetical similar CDF experiments would have produced a more significant 
region purely by fluctuations of the SM background. The results of all three 
global-search algorithms have not yet shown evidence of new physics. 
 
\subsection{\boldmath Search for Large Extra Dimensions in $\g\met\;$ \label{sec:LED}}
\begin{figure}
\begin{tabular}{cc}
\includegraphics[width=0.4\textwidth]{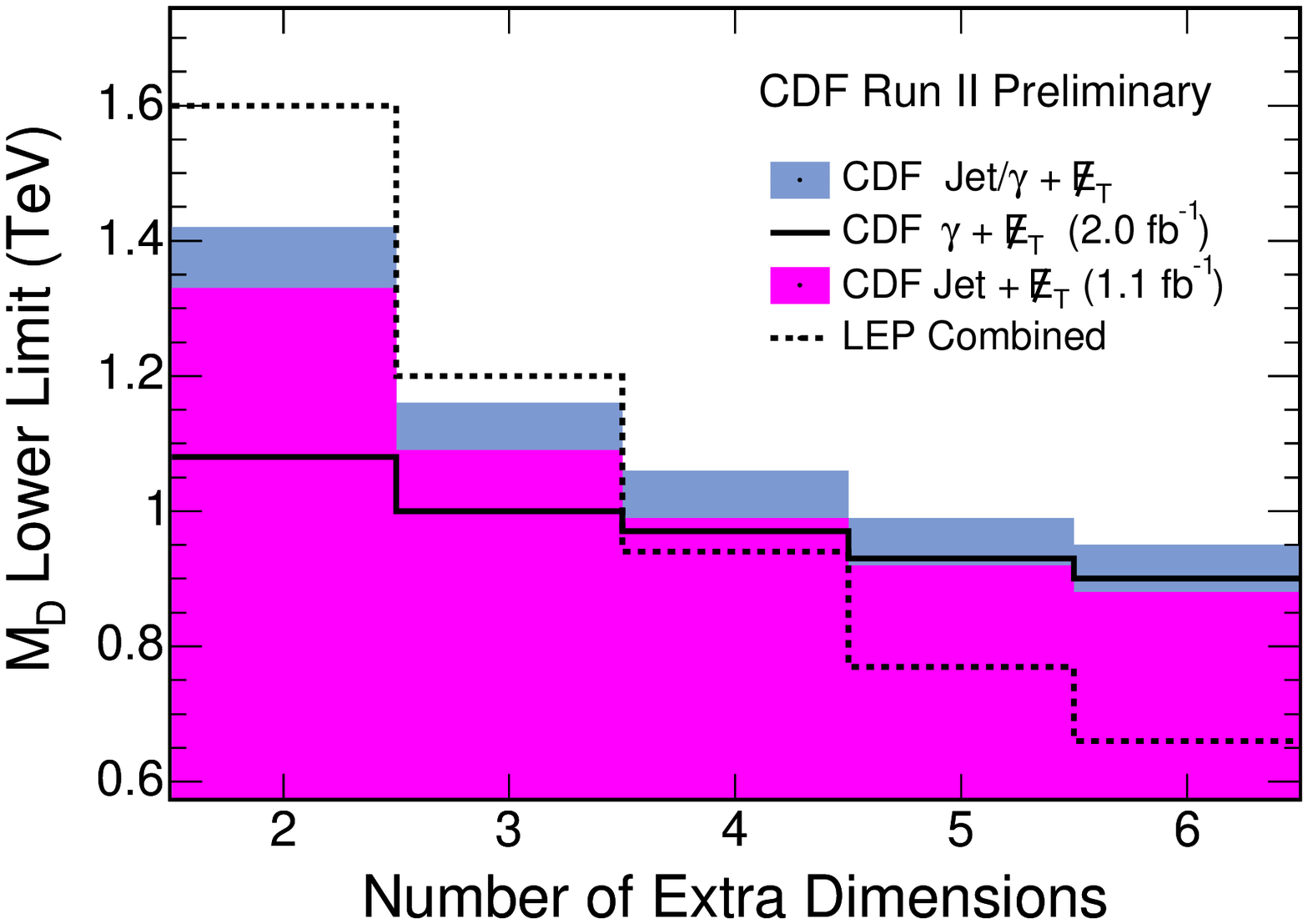} &   
\includegraphics[width=0.4\textwidth]{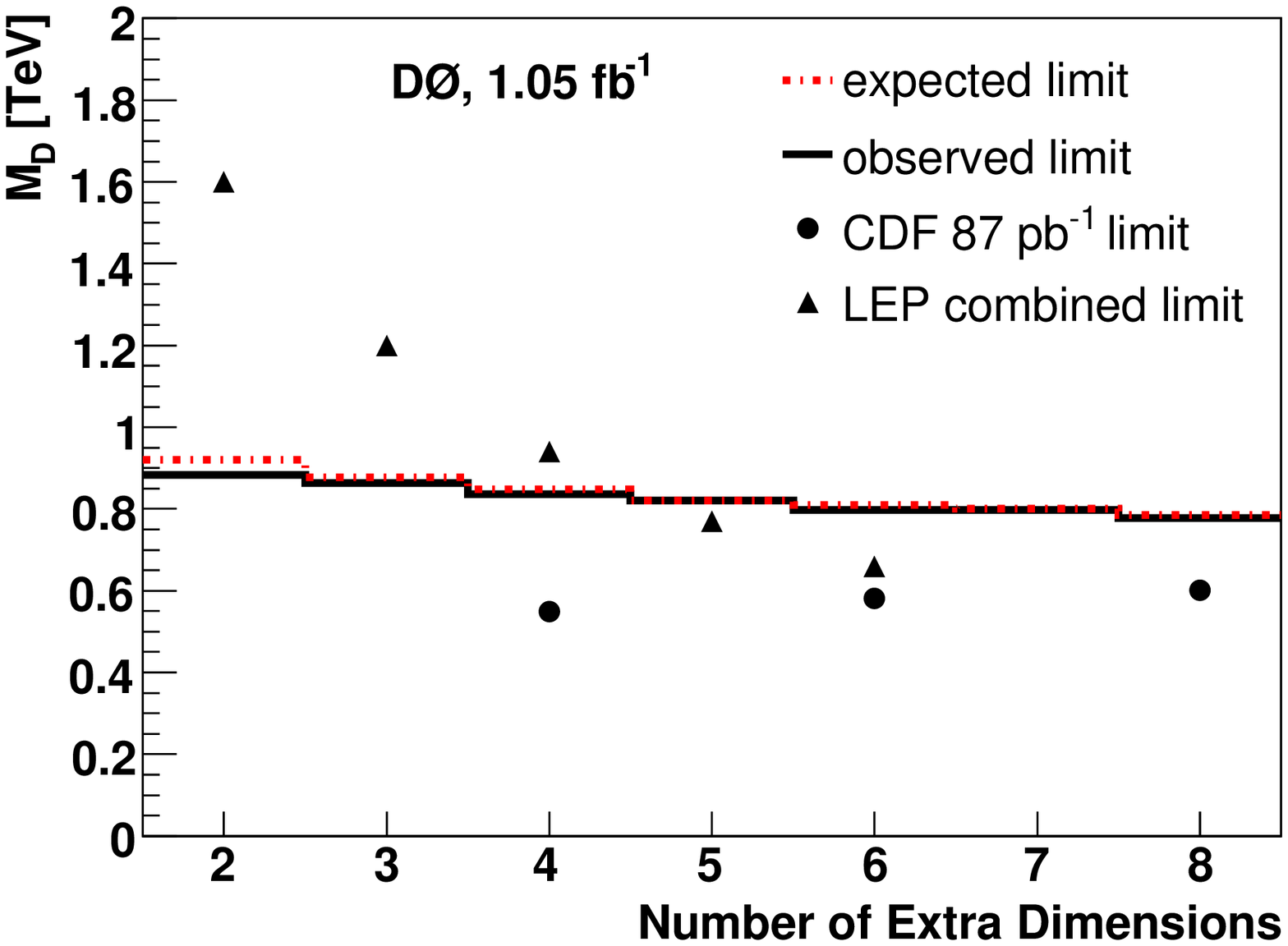} \\
\end{tabular} 
	\caption{\label{fig:LED}
	Search for large extra dimensions in $\g\met\;$: the 95\% CL lower 
	limits on the fundamental Plank mass $M_D$ vs. number of extra 
	dimensions from CDF (left) and D0 (right), compared with the limits 
	set by the LEP experiments.  
	A combined limit from the two LED searches at CDF, using $\g\met\;$ and
	monojet+$\met\;$ final states, is also shown.
	}
\end{figure}

\unboldmath
The CDF and D0 collaborations have looked for indications of large extra 
dimensions (LED)~\cite{ArkaniHamed:1998rs} in 2.0~\fbarn\ and 1.1~\fbarn\ of data, respectively~\cite{Krutelyov:2008nb,Abazov:2008kp}. In the LED model, the 
production $q\bar{q}\rightarrow \g G$ gives an exclusive $\g\met\;$ final state
 where the $\met\;$ arises from the massive and non-interacting graviton. The 
analyses require one central photon with $\et > 90~\gev$ and 
$\met\; > 50/70~\gev$ for CDF/D0. Events with extra high \pt\ tracks or jets 
are removed. The exclusive $\g\met\;$ final state suffers from large amount of 
cosmic rays and beam halos and the analysis would have been impossible if 
an effective rejection was not applied. The CDF analysis requires the photon 
to be in time with a $p\bar{p}$ collision and uses topological variables to 
separate signal from non-collision background, such as track multiplicity, 
angular separation between the photon and the closest hit in the muon chamber, 
and energy deposited in the calorimeters. 
The D0 analysis utilizes the transverse and the unique longitudinal 
segmentation of the electromagnetic (EM) calorimeter. The photon trajectory is 
reconstructed by fitting one measurement in the preshower detector and four in 
the EM calorimeter to a straight line (EM pointing algorithm). The $z$ 
position and the transverse impact parameter of the photon, at the point of 
closest approach with respect to the beam line, are required to be within 
10~cm and 4~cm of a $p\bar{p}$ interaction vertex, respectively\footnote{The resolution of 
both the $z$ position and the transverse impact parameter is about 2~cm.}.  
The distribution of the transverse impact parameter is further used to 
estimate the amount of remaining non-collision background. 
After all selections, the dominant background in both analyses is SM 
$Z\gamma \rightarrow \nu\nu\gamma$ production. Both analyses have not found 
significant excess in data: 
40 observed vs. $46.3\pm3.0$ expected (CDF) and 
29 observed vs. $22.4\pm2.5$ expected (D0). 
Lower limits on the fundamental Plank scale, $M_D$, are set at 95\% confidence 
level (CL) as a function of the number of extra dimensions, $n$ (see Figure~\ref{fig:LED}). 
For $n=4$, the lower limit on $M_D$ is 970~\gev\ for CDF and 836~\gev\ 
for D0. The Tevatron results supersedes the LEP limits~\cite{pdg} when $n > 3$ 
for CDF and when $n > 4$ for D0. 

\subsection{\boldmath Search for High-mass $ee$ and $\gamma\gamma$ Resonances} 
\begin{figure}
\begin{tabular}{cc}
\includegraphics[width=0.3\textwidth]{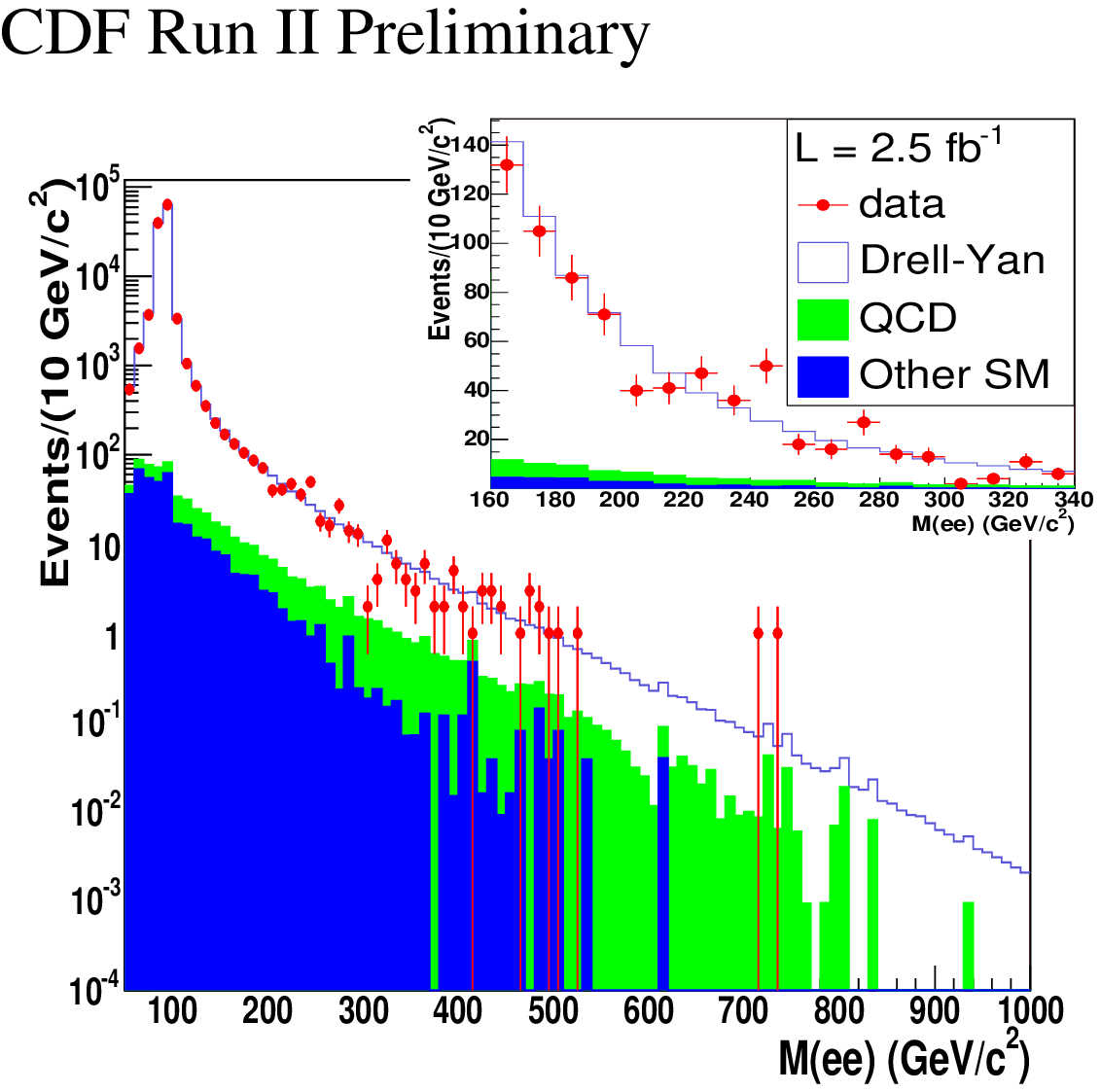} &   
\includegraphics[width=0.4\textwidth]{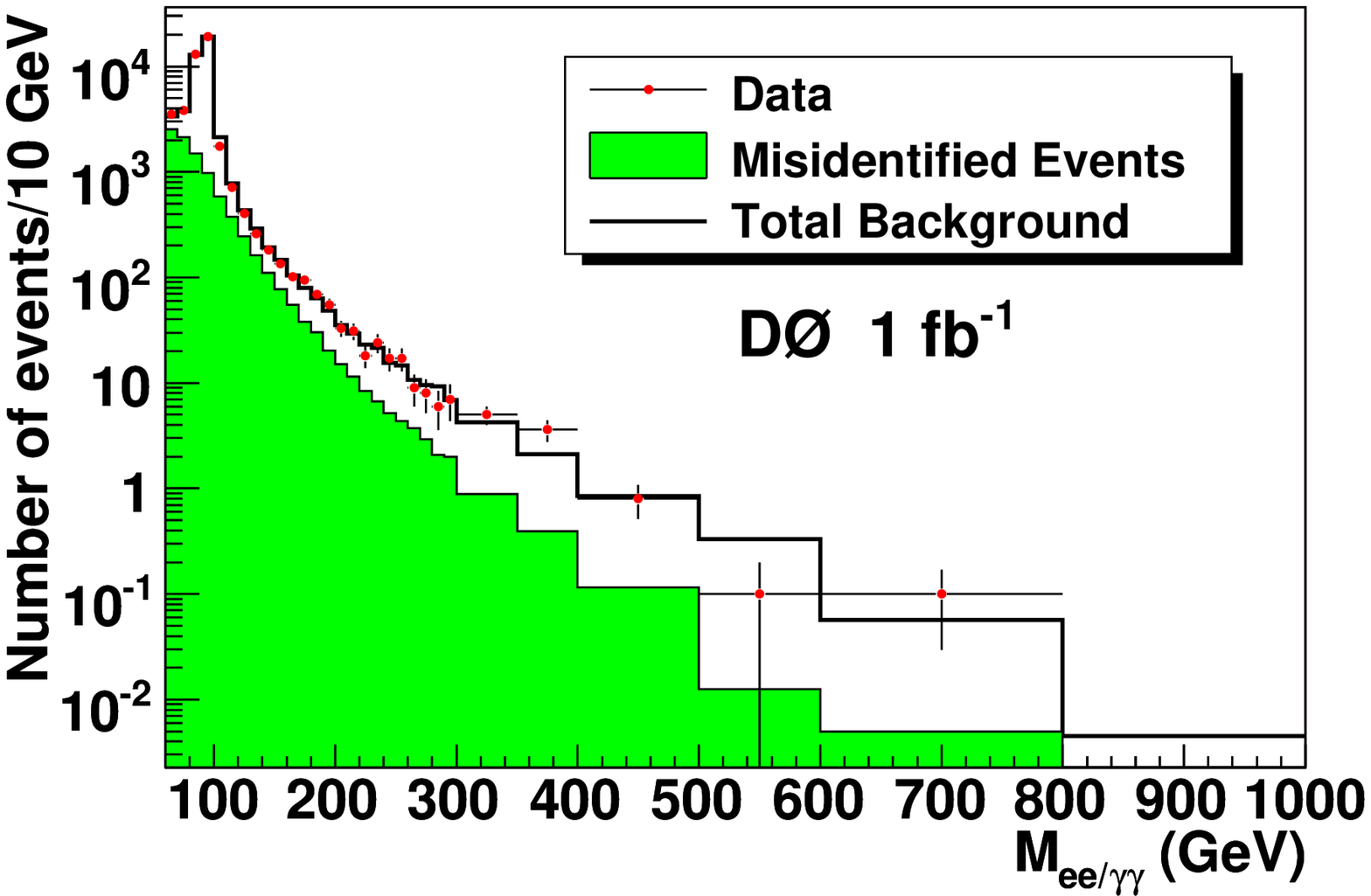} \\
\end{tabular} 
	\caption{\label{fig:highmass}
	Search for high-mass $ee$, $\gamma\gamma$ resonances: the 
	$M_{ee}$ spectrum from CDF (left) and $M_{ee,\g\g}$ spectrum 
	from D0 (right), observed (markers) and background prediction 
	(filled histograms). 
	The CDF data have a $3.8~\sigma$ excess for the mass window 
	$ 228<M_{ee}<250~\gevcsq$.
	}
\begin{tabular}{cc}
\includegraphics[width=0.3\textwidth]{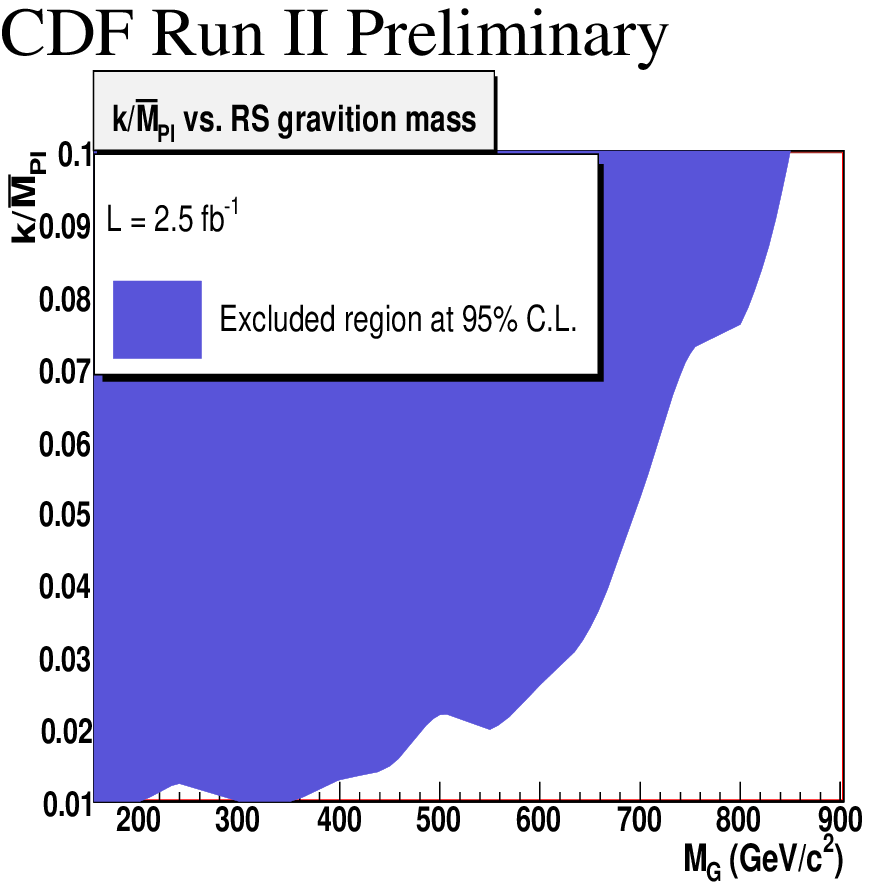} &   
\includegraphics[width=0.3\textwidth]{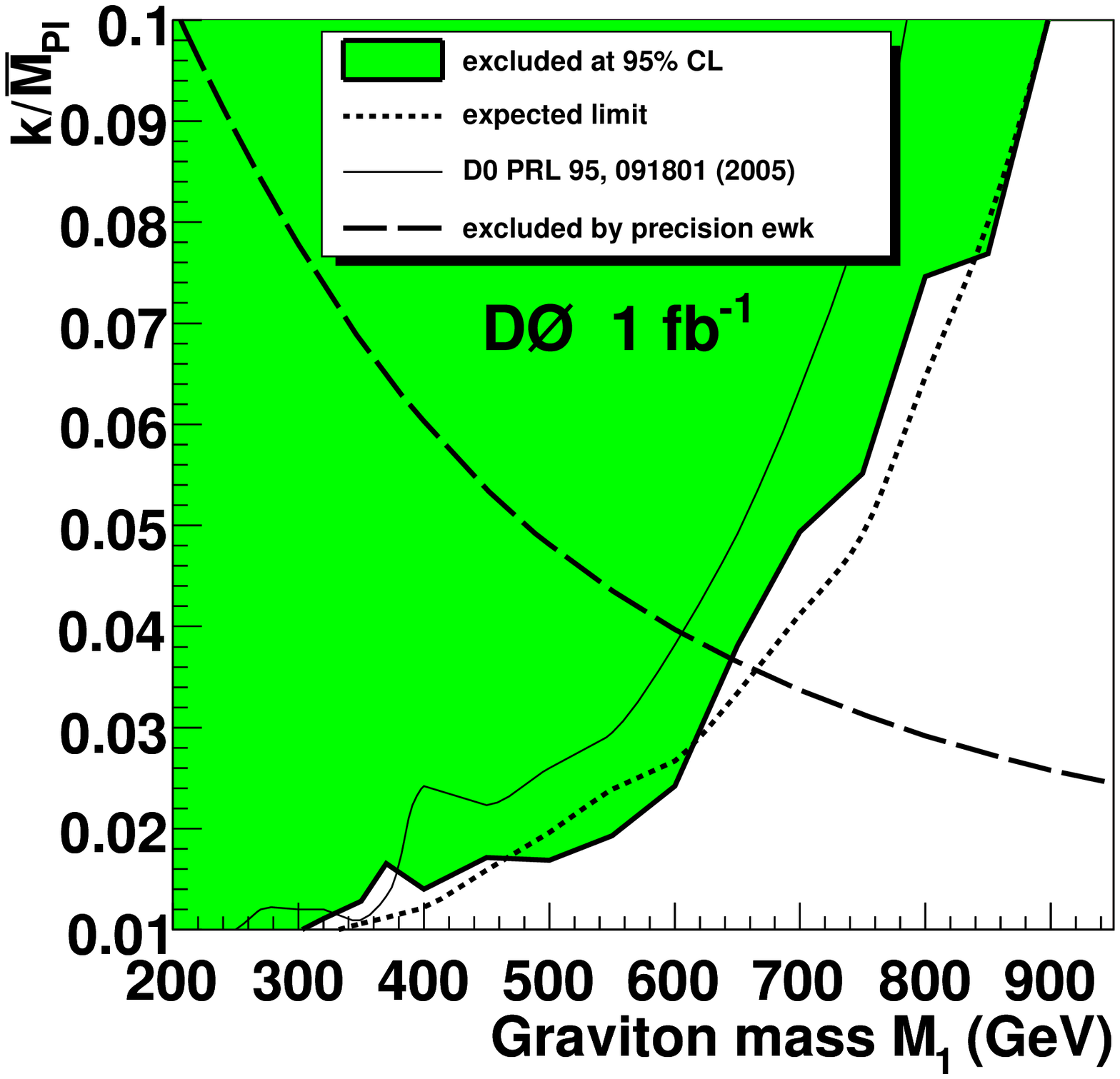} \\
\end{tabular} 
	\caption{\label{fig:RS}
	The excluded regions of RS graviton mass 
	with respect to $k/\bar{M}_{Pl}$ from CDF (left) and D0 (right). 
	}
\end{figure}

\unboldmath
Many extensions of the standard model have predicted new particles 
which decay to a lepton-lepton or photon-photon pair, such as Randall-Sundrum 
(RS) graviton~\cite{Randall:1999ee} and $\zp$ from the $E_6$ 
model~\cite{Hewett:1988xc}. The CDF and the D0 collaborations have searched 
for high-mass resonances in the $ee$ and $ee/\g\g$ final states, using 2.5 
and 1.0~\fbarn\ of data, respectively~\cite{Abazov:2007ra}. 
The CDF analysis requires two electrons 
in the central-central or central-forward\footnote{The forward electrons have 
detector pseudo-rapidity $1.2 <\left|\eta^\mathrm{det}\right| < 2.0$.} region 
while the D0 analysis requires two electromagnetic (EM) objects\footnote{The 
EM objects have no requirements on tracks 
and include both electrons and photons.} in the central-central region;  
the CDF electrons and the D0 EM objects must have $\et > 25~\gev$ each. 
Figure~\ref{fig:highmass} shows the $M_{ee}$ and $M_{ee/\g\g}$ spectra from 
CDF and D0, individually. The dominant background is SM Drell-Yan 
production (and also diphoton production for D0). The D0 data are consistent 
with the background prediction while the CDF data have a $3.8~\sigma$ excess 
for the mass window $228<M_{ee}<250~\gevcsq$. The probability (or the 
$p$-value) to observe such an excess anywhere in the search window 
$150 < M_{ee} < 1000~\gevcsq$ is 0.6\%. Without significant excess in both 
analyses, CDF and D0 set limits on the mass of RS graviton with 
respect to the coupling between the RS graviton and the SM particles, 
$k/\bar{M}_{Pl}$\footnote{Here, $k$ is the warp factor which gives the 
curvature of extra dimension in the RS model and $\bar{M}_{Pl}$ is the reduced 
Plank scale.} (see Figure~\ref{fig:RS}). For $k/\bar{M}_{Pl}=0.1$, masses below
 850 (CDF) and 900 (D0)~\gevcsq\ are excluded at 95\% CL. 
CDF also sets the world's best lower mass 
limits for $\zp$ boson with SM coupling and those predicted by the $E_6$ 
model (see Table~\ref{t:Zprime}).

\begin{table}
\begin{center}
\caption{Expected and observed lower mass limits for $\zp$ boson with SM coupling and those predicted by the $E_6$ model. These limits have been set by CDF 
using the results of search for high-mass $ee$ resonances.}
\begin{tabular}{|c|c|c|c|c|c|c|c|}
\hline 
 & $\zp_\mathrm{SM}$ & $\zp_{\Psi}$ & $\zp_{\chi}$ & $\zp_{\eta}$ & $\zp_{I}$ 
 &  $\zp_{sq}$ & $\zp_N$
\\ 
 \hline
 Exp. Limit (\gevcsq) & 965 & 849 & 860 & 932 & 757 & 791 & 834 \\
 Obs. Limit (\gevcsq) & 966 & 853 & 864 & 933 & 737 & 800 & 840 \\
\hline
\end{tabular}
\label{t:Zprime}
\end{center}
\end{table}

\subsection{\boldmath Search for $W^{\prime} \rightarrow e\bar{\nu}_e$}
\unboldmath
Additional charged gauge boson, $W^{\prime}$, has been introduced by 
several new physics models, such as left-right symmetric model~\cite{LRModel} 
and the $E_6$ model~\cite{Wprime}. The D0 collaboration has searched 
for a $W^{\prime}$ decaying to an electron and a neutrino using 1~\fbarn\ 
of data~\cite{D0Wprime}. Events are required to have a central electron with 
$\et > 30~\gev$ and $\met\; > 30~\gev$. Clean-up cuts are applied to reduce 
mis-measured $\met\;$. Data with transverse mass\footnote{The 
transverse mass is defined as \(m_T =\sqrt{2E_T^{el}\met\;\left(1-cos\Delta\phi\right)}\), where $E_T^{el}$ is the transverse 
energy of electron and $\Delta\phi$ is the azimuthal angle between 
the electron and missing energy.} $m_T < 30~\gevcsq$ 
and $60 < m_T < 140~\gevcsq$ are used to obtain the normalizations of QCD 
multi-jet and SM $W\rightarrow e\bar{\nu}_e$ backgrounds, separately. 
There is no excess in the search window $140< m_T < 1000~\gevcsq$ 
(see Figure~\ref{fig:WprimeDijet}). The 
shape of $m_T$ distribution serves as a discriminant to separate the exotic 
signal from the SM background when setting the lower mass limit on 
$W^{\prime}$. Using the 
Altarelli reference model~\cite{WprimeRef} where SM couplings are assumed, 
 $W^{\prime}$ with mass below 1~TeV$/c^{2}$ is excluded at 95\% CL. This limit 
is currently the world's best limit.


\subsection{Search for High-mass Dijet Resonances}
\begin{figure}
\begin{tabular}{cc}
\includegraphics[width=0.3\textwidth]{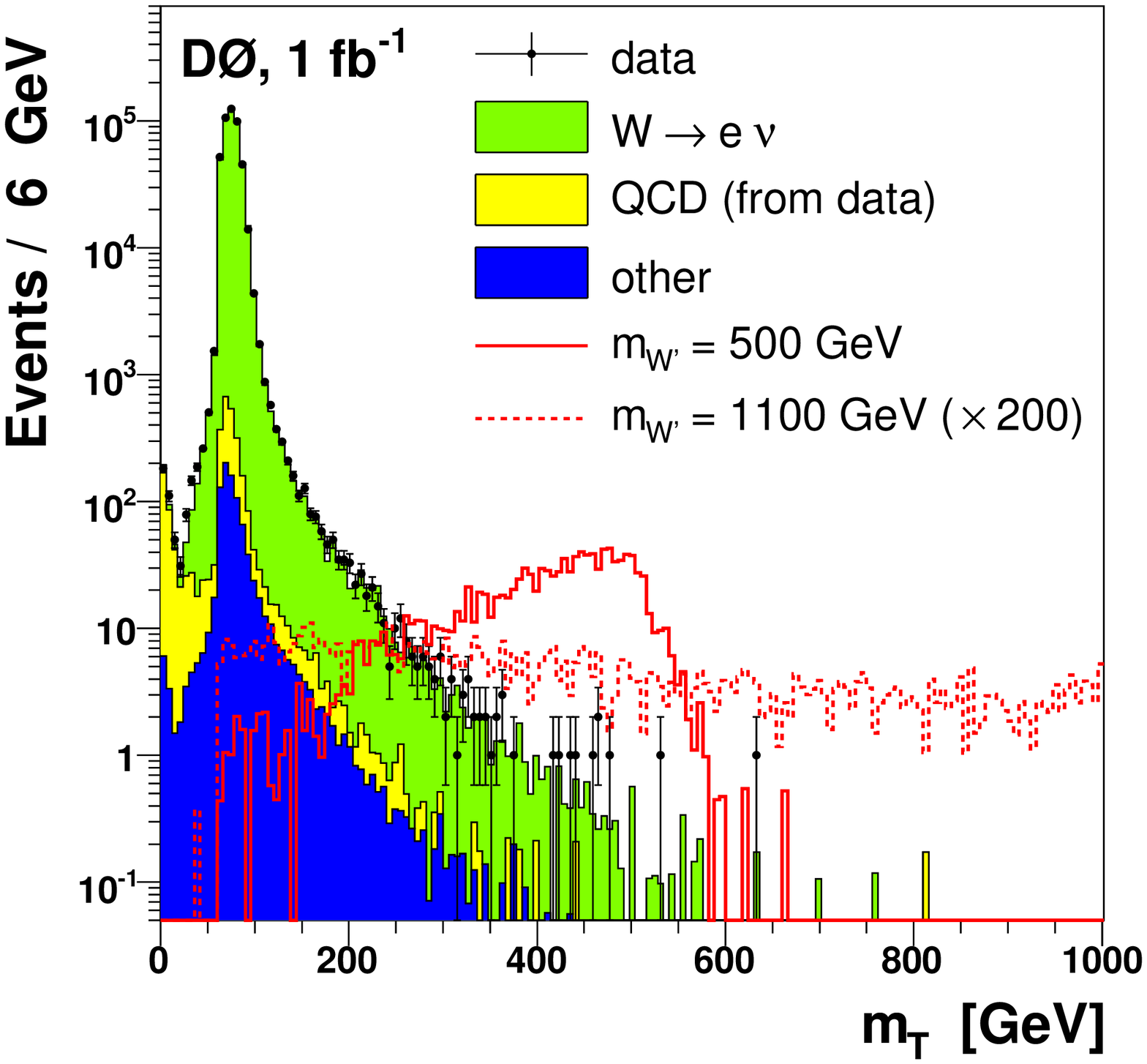} &   
\includegraphics[width=0.4\textwidth]{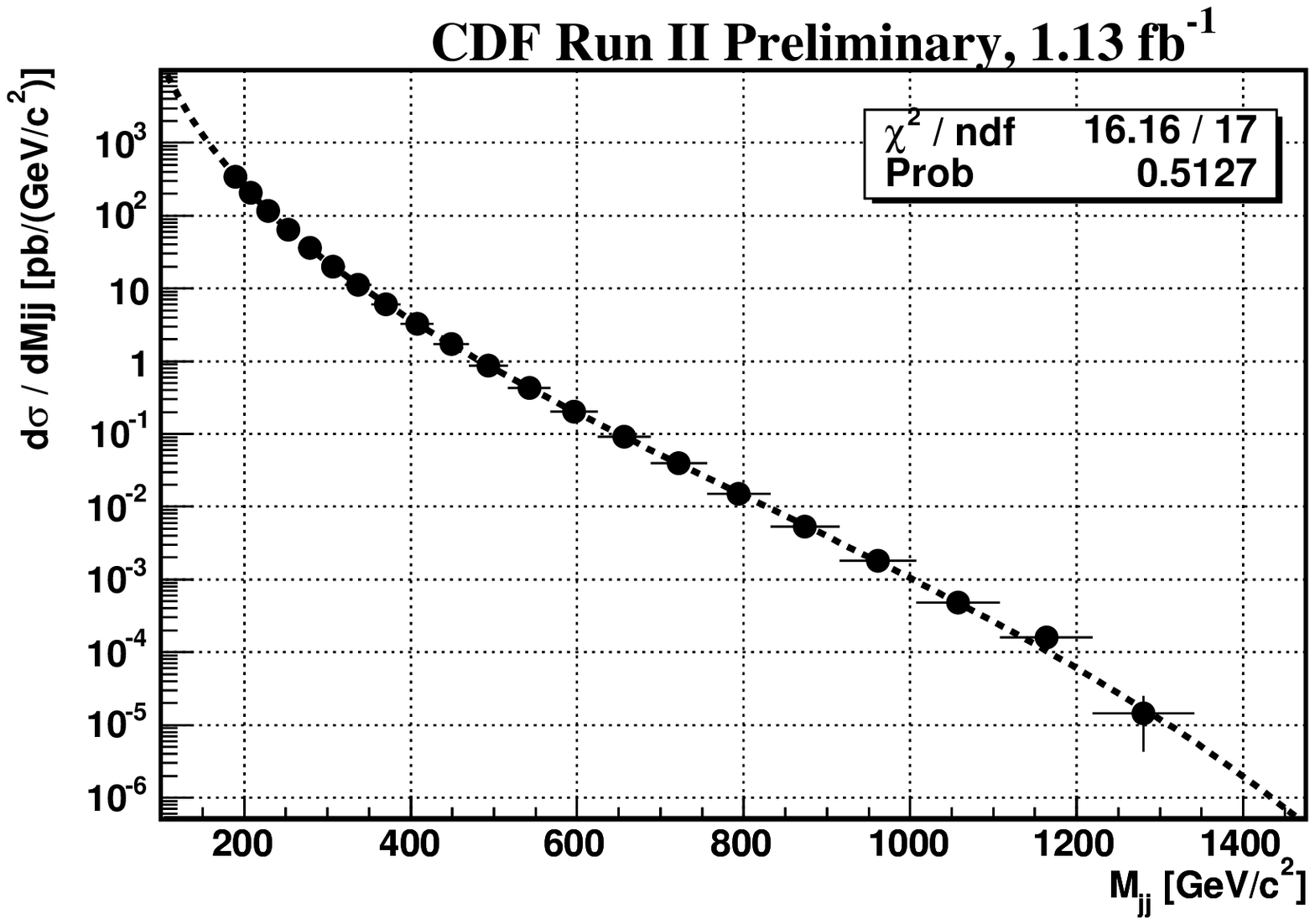} \\
\end{tabular} 
	\caption{\label{fig:WprimeDijet}
	The left figure shows the distribution of transverse mass, $m_T$, 
	observed (markers) and background prediction 
	(filled histograms) from the D0 search for 
	$W^{\prime}\rightarrow e\bar{\nu}_e$. 
	The expected $m_T$ distributions for $W^{\prime}$ with masses 
	at 500~\gevcsq\ and 1100~\gevcsq\ are also shown. 
 	The right figure shows 
	the $M_{jj}$ spectrum observed in the CDF data. The fit function,  
	\(\frac{d\sigma}{dm} = p_0(1-x)^{p_1}\left/x^{p_2+p_3\log x},\right.\)
	where \(x= m\left/\sqrt{s},\right.\) describes the data well.}
\end{figure}

New particles which decay into two energetic partons (quarks and gluons) are 
expected to produce a resonant structure in the dijet mass spectrum. 
Such new particles include excited quarks ($q^*\rightarrow qg$)~\cite{Baur:1987ga}, axigluons ($A\rightarrow q\bar{q}$)~\cite{Bagger:1987fz}, color octet 
techni-$\rho$ ($\rho_T \rightarrow q\bar{q}, gg$)~\cite{Lane:1991qh}, 
$W^{\prime}$ ($W^{\prime}\rightarrow q\bar{q}^{\prime}$) , $Z^{\prime}$ ($Z^{\prime}\rightarrow q\bar{q}$), diquarks in the string-inspired $E_6$ model 
[$D(D^c)\rightarrow (qq)\bar{q}\bar{q}$]~\cite{Hewett:1988xc}, and Randall-Sundrum graviton ($G\rightarrow q\bar{q}, gg$)~\cite{Randall:1999ee,Bijnens:2001gh}. The CDF collaboration has performed a search for high-mass dijet resonances 
in 1.1~\fbarn\ of data. Events are required to have two central jets with 
invariant mass $M_{jj}>180~\gevcsq$ where the jet energy is corrected to the 
hadron level, and events must not have significant $\met\;$. The background is 
completely dominated by the QCD dijet production. The measured $M_{jj}$ 
spectrum is fit to a smooth function motivated by predictions of 
{\sc PYTHIA} and {\sc HERWIG} MC and calculations by the {\sc NLOJET++} program
 (see Figure~\ref{fig:WprimeDijet}). No excess of data above the fit is 
observed. This analysis has set the world's best limits on excited quarks, 
axigluon and coloron, color octet techni-$\rho$, and $E_6$ diquarks, and 
excluded the mass regions $260 < M(q^*) < 870~\gevcsq$, 
$260 < M_A < 1250~\gevcsq$, $260 < M(\rho_T) < 1100~\gevcsq$, 
and $ 260 < M(D,D^c) < 630~\gevcsq$, at 95\% CL,respectively.

\subsection{\boldmath Search for New Physics in Exclusive $jj\met\;\;$ and the 
Leptoquark Interpretation} 
\begin{figure}
\begin{tabular}{cc}
\includegraphics[width=0.4\textwidth]{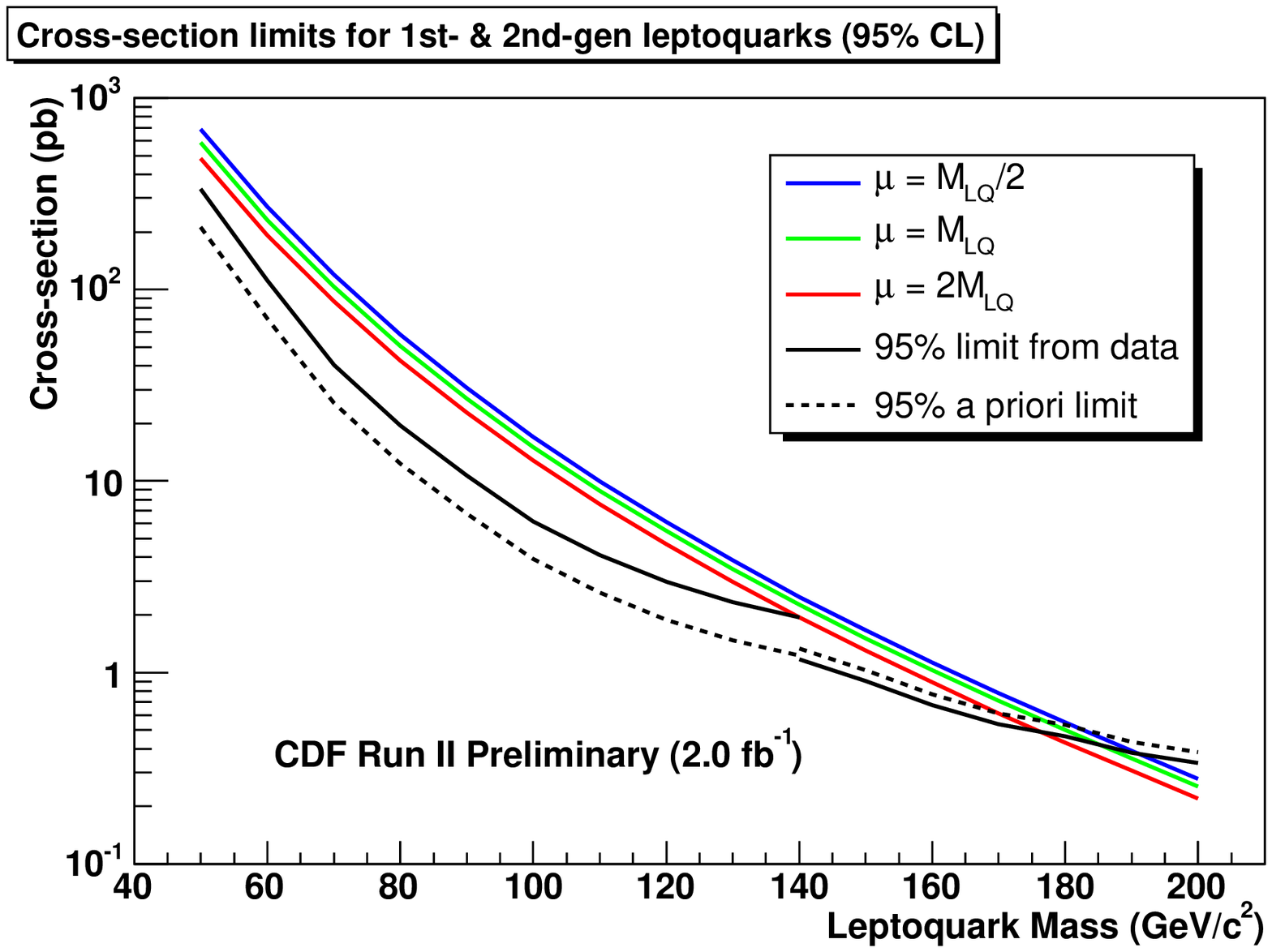} &   
\includegraphics[width=0.4\textwidth]{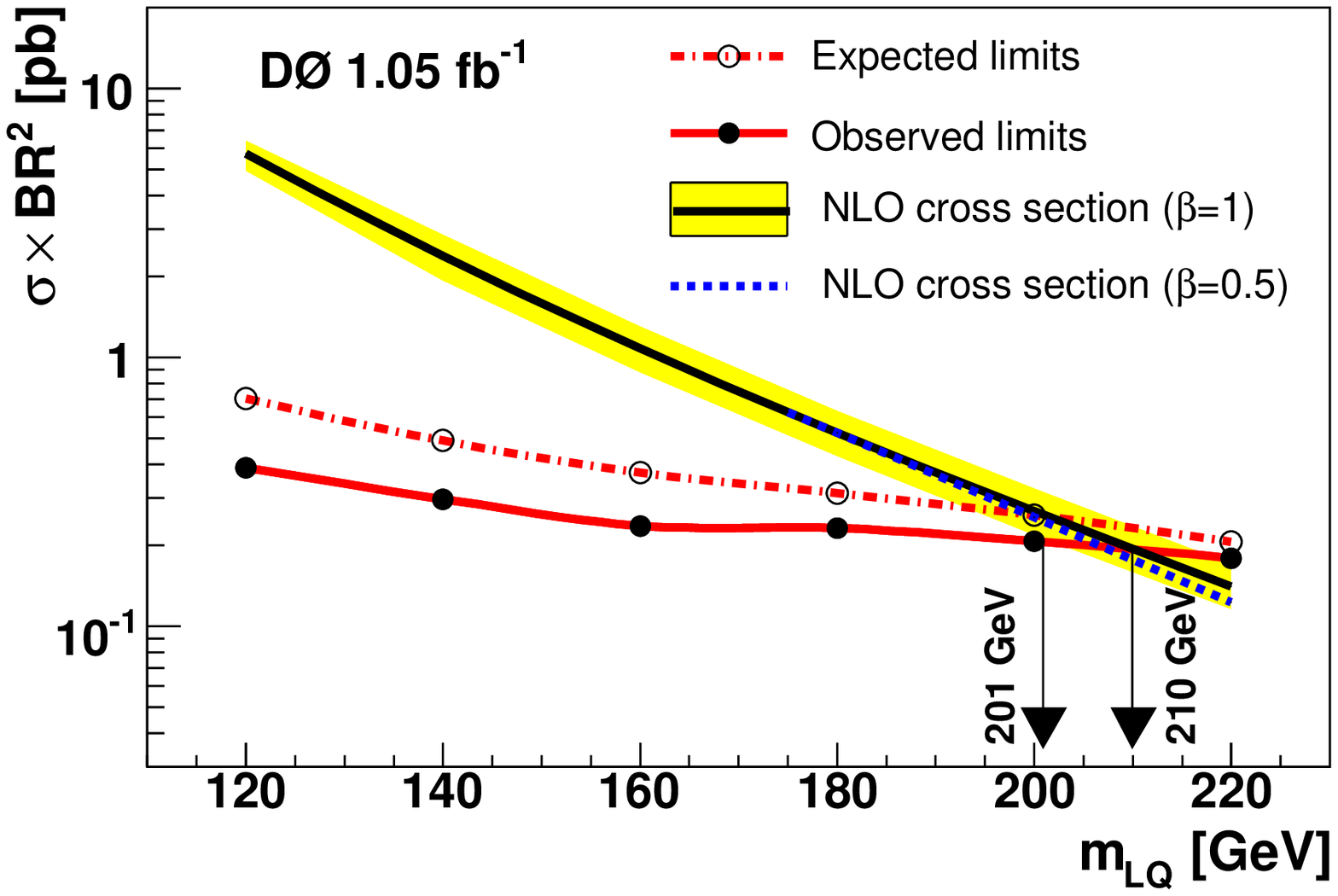} \\
\end{tabular} 
	\caption{\label{fig:LQ}
	The observed and expected cross-section limits 
	on the pair production of scalar leptoquarks
	for the first and the second generations 
	from CDF (left), and for the third generation from D0 (right). 
	Theoretical predictions for different renormalization scales are 
	also shown. The discontinuity of CDF limits is due to the use of 
	low/high kinematic regions for mass below/above 
	140~\gevcsq\ to obtain the best sensitivity. 
	}
\end{figure}

\unboldmath
The signature with exclusive dijet and large $\met\;$ has been predicted by 
leptoquarks~\cite{pdgLQ}, SUSY~\cite{Goldberg:1983nd}, Universal Extra 
Dimensions with conservation of the momentum in the volume of the extra 
dimensions~\cite{Servant:2002aq}, and Little Higgs with T-parity 
conservation~\cite{Cheng:2003ju}. The CDF collaboration has extended its 
previous monojet + $\met\;$ search to the $jj\met\;$ channel using 
2.0~\fbarn\ of data. Events are required to have exactly two jets with 
$\et > 30~\gev$ and $\left|\eta_\mathrm{det}\right| < 2.4$, no extra jets with 
$\et > 15~\gev$. Events containing EM objects and isolated tracks are removed. 
In order to be sensitive to different scenarios of new physics, 
two kinematic regions are defined. 
The ``low kinematic region'' must have $\met\; > 80~\gev$ and scalar sum \et\ 
of two jets $E_T^{j1} + E_T^{j2}$ $> 125~\gev$, while the 
``high kinematic region'' must have $\met\; > 100~\gev$ and 
$E_T^{j1} + E_T^{j2}$ $> 225~\gev$. The dominant backgrounds are SM productions
 of $W$ + jets $\rightarrow \ell\nu$ + jets with a missing lepton 
and $Z$ + jets $\rightarrow \nu\nu$ + jets. Data agree well with the background
 prediction: 2506 observed vs. $2312 \pm 140$ expected (low kinematic) and 
186 observed vs. $196\pm 29$ expected (high kinematic). The results are turned 
to limits on the masses of the first ($LQ_1$) and the second generation scalar leptoquarks ($LQ_2$). The leptoquarks are pair produced 
and both charge $1/3$ and charge $2/3$ leptoquarks are included. 
The $LQ_1$ and $LQ_2$ are assumed to decay to $\nu_{\ell}q$ with a unity 
coupling. When the renormalization scale $\mu$ is set to be twice of the 
leptoquark mass, the lower mass limits on $LQ_1$ and $LQ_2$ are 177~\gevcsq\ 
(see Figure~\ref{fig:LQ}). These are currently the world's best limits. 

\subsection{\boldmath Search for Third Generation Leptoquark in $\tau^+\tau^- b\bar{b}$} 
\unboldmath
Leptoquarks are predicted in many models to explain the observed symmetry  
between leptons and quarks, such as Technicolor~\cite{Dimopoulos:1979es}, grand
 unification~\cite{Pati:1974yy}, superstrings~\cite{Hewett:1988xc}, and 
quark-lepton compositeness~\cite{Schrempp:1984nj}.
The D0 collaboration has looked in 1.1~\fbarn\ of data for pair production of 
third generation scalar leptoquarks\footnote{Given the null evidence of flavor 
changing neutral current, leptoquarks of each generation are expected to 
couple only to fermions of the same generation.} ($LQ_3$) in the 
$\tau^+\tau^- b\bar{b}$ final state~\cite{D0LQ}. 
Both charge $2/3$ and charge $4/3$ leptoquarks are included.
 Events must have 
a muon with $\pt > 15~\gevc$ and $\left|\eta_\mathrm{det}\right| < 2.0$, 
a hadronic $\tau$ with visible $\pt>15~\gevc$, at least two jets with 
$\et>25,20~\gev$ and $\left|\eta_\mathrm{det}\right| < 2.5$ and at least 
one of the jets must be ``$b$-tagged'' by a neural network 
algorithm~\cite{Scanlon:2006wc}. A maximum requirement on the variable related 
to the $W$ boson mass\footnote{The $m^*$ is defined as \(\sqrt{2E^{\mu}E^{\nu}
\left(1-cos\Delta\phi\right)}\), where the estimated neutrino 
energy is \(E^{\nu} = \met\;\times\left(E^{\mu}/\pt^{\mu}\right)\) and 
$\Delta\phi$ is the azimuthal angle between the muon and missing energy.}, 
$m^* < 60~\gevcsq$, is applied to suppress SM background which contains a $W$ 
($t\bar{t}$ and $W$ + jets). The dominant 
backgrounds after all selections are $Z$ + jets and $t\bar{t}$ productions. 
No excess has been observed in either the exactly one $b$-tag events 
(15 observed vs. $19.6\pm2.5$ expected) or the $\geq 2$ $b$-tag events 
(1 observed vs. $4.8\pm1.0$ expected). The variable $S_T$, which is the scalar 
sum \pt\ of the muon, hadronic tau, and two highest \pt\ jets, is expected 
to be higher for the $LQ_3$ signal than for the SM background. The distribution
 of $S_T$ is used as a discriminant to set lower mass limits on $LQ_3$. 
The 95\% CL lower mass limit on scalar $LQ_3$ is 210~\gevcsq\ when 
the coupling constant\footnote{The charge 2/3 $LQ_3$ decays to $\tau^+ b$ with 
coupling constant $\beta$ and to $\bar{\nu}_{\tau}t$ with coupling $(1-\beta)$.} $\beta$ is 1 and 207~\gevcsq\ when $\beta$ is 0.5. Both limits are the 
world's best limits.

\subsection{Search for Maximal Flavor Violation in Same-sign Tops}
In the model of maximal flavor violation (MxFV)~\cite{BarShalom:2007pw}, there 
is at least one new scalar $\Phi_{FV}\equiv\left(\eta^+,\eta^0\right)$ which 
couples to quarks via $\Phi_{FV}q_i q_j \propto \xi_{ij}$, where 
$\xi_{i3},\xi_{3i}\sim V_{tb}$ for $i=1,2$ and $\xi_{33}\sim V_{td}$ and 
$V$ is the CKM matrix~\cite{CKM}. When 
$\xi\equiv\xi_{31}=\xi_{13}\sim {\cal O}(1) \gg \xi_{23},\xi_{32}\gg\xi_{33}$, 
$\eta^0$ decays half of the time to $t+\bar{u}$ and half the time to 
$\bar{t}+u$. If the charged scalar $\eta^+$ is too heavy to access at Tevatron 
or LHC and the neutral scalar $\eta^0$ is light, a striking signature with 
same-sign top quark pairs may be produced through 
\(ug\rightarrow t\eta^0\rightarrow tt\bar{u} + h.c.\), 
\(u\bar{u}\rightarrow \eta^0\eta^0 \rightarrow tt\bar{u}\bar{u} + h.c.\), 
and \(uu\rightarrow tt + h.c.\), where the last process comes from 
$t$-channel $\eta^0$ exchange~\cite{BarShalom:2008fq}.
 The CDF collaboration has searched for same-sign tops predicted by MxFV in 
2.0~\fbarn\ of data. Events are required to have a pair of same-sign leptons 
(electron or muon) with $\pt > 20~\gevc$, $\geq 1$ jet $b$-tagged by 
a jet probability tagging algorithm~\cite{JetProb}, and 
$\met\; > 20~\gev$. The dataset has strong sensitivity to this signature: 
if $m_{\eta^0}\sim 200~\gevcsq$ and $\xi~\sim 1$, $\sim$ 11 MxFV events are 
expected over a background of $2.9\pm 1.8$ events. There are 3 events observed 
in data, which is consistent with the background prediction, and 
95\% CL limits are set on $m_{\eta^0}$ and the coupling $\xi$.
Figure~\ref{fig:MxFVTC} shows the allowed mass of $\eta^0$ with respect to 
$\xi$. At $m_{\eta^0} = 200~\gevcsq$, $\xi < 0.85$.

\subsection{\boldmath Search for Technicolor Particles $\rho_T^0$ and $\rho_T^{-}$}
\begin{figure}
\begin{tabular}{cc}
\includegraphics[width=0.35\textwidth]{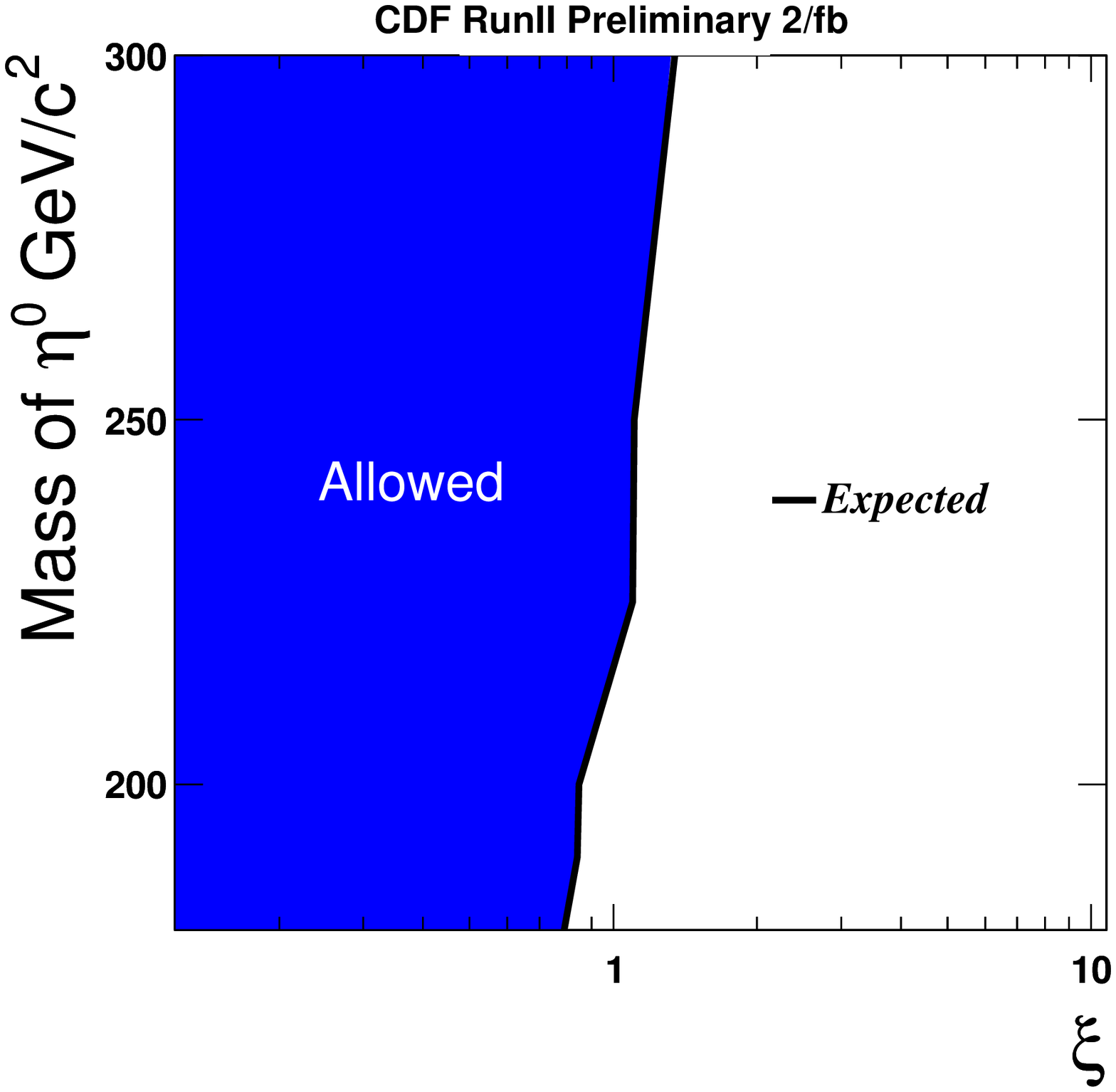} &   
\includegraphics[width=0.34\textwidth]{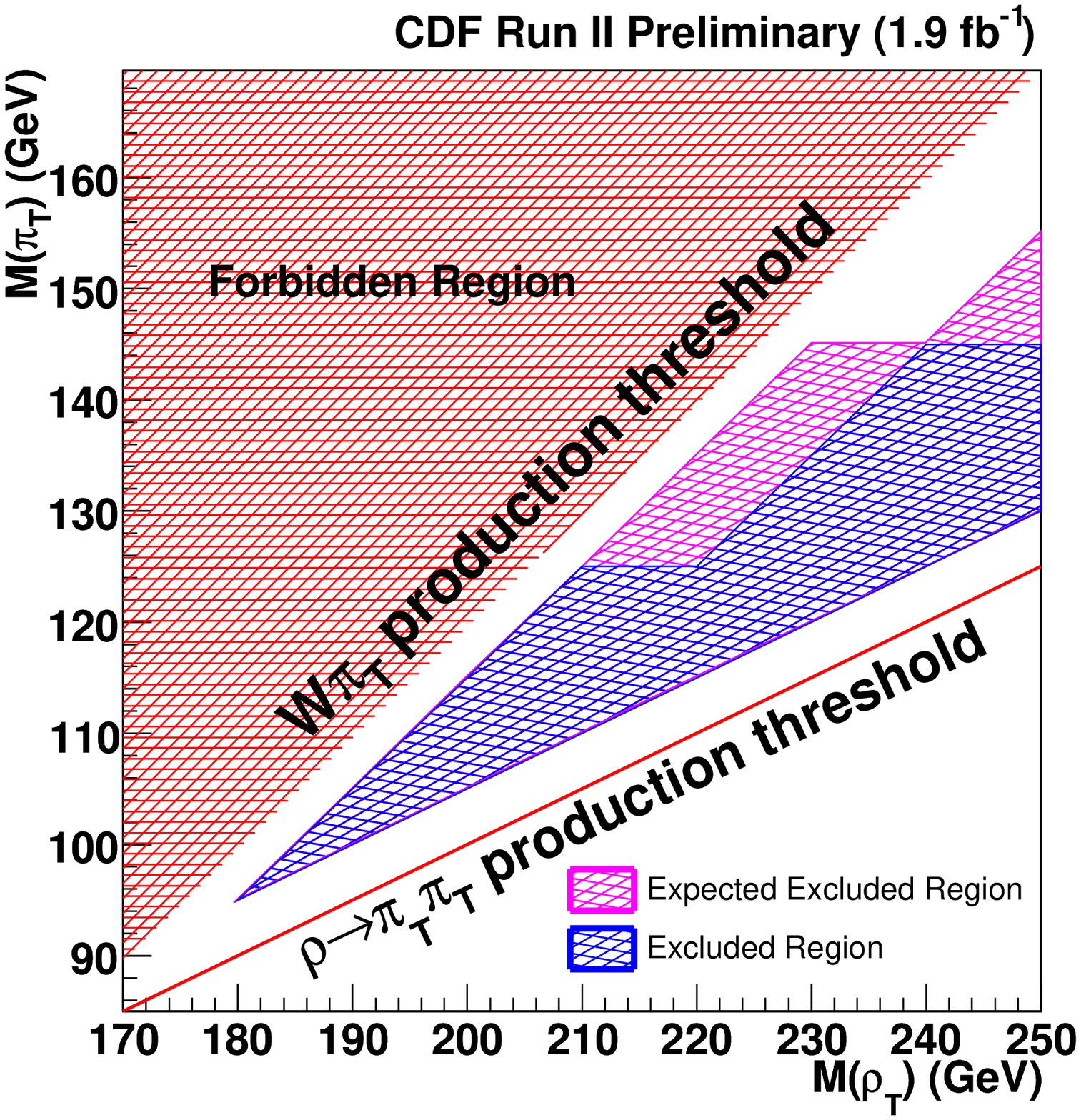} \\
\end{tabular} 
	\caption{\label{fig:MxFVTC}
	The observed allowed region in the $m_{\eta^0}-\xi$ plane (left) 
	and the observed and expected excluded regions in the 
	$M(\pi_T)-M(\rho_T)$ plane. Both analyses are performed by the 
	CDF collaboration.
	}
\end{figure}

\unboldmath
Technicolor~\cite{Dimopoulos:1979es} provides an alternative to explain the electroweak symmetry breaking, in addition to the Higgs mechanism. Both 
mechanisms predict new particles which could be produced in association with a 
$W$ boson. Using 1.9~\fbarn\ of data,
the CDF collaboration has extended its search for SM Higgs, 
$p\bar{p}\rightarrow WH_\mathrm{SM}\rightarrow Wb\bar{b}$, 
to a search for technicolor rhos and pions via the decay chain: 
\(p\bar{p} \rightarrow \rho_T^{-} \rightarrow W^{-}\pi_T^0 \rightarrow 
\ell\nu_{\ell} b\bar{b}\) and \(p\bar{p} \rightarrow \rho_T^{0} \rightarrow 
W^{-}\pi_T^{+} \rightarrow \ell\nu_{\ell} c\bar{b}, 
\ell\nu_{\ell} u\bar{b}\). 
Events must have a central electron or muon with $\pt>20~\gevc$, 
exactly two jets with $\et > 20~\gev$ and $\left|\eta_\mathrm{det}\right| < 2.0$, and $\met\; > 20~\gev$. Three types of $b$-tagging requirements are applied:
1. exactly one $b-$tagged by the tight SECVTX and a neural network 
algorithm~\cite{SECVTX,NN}, 
2. two $b$-tagged, both by the tight SECVTX algorithm, 
3. two $b$-tagged, one by the tight SECVTX, and one by a jet probability 
tagging algorithm~\cite{JetProb}. These three classes of events have different 
signal purities and are analyzed separately. Data agree with background 
prediction in all categories: 805 observed vs. $810\pm159$ expected (class 1), 
83 observed vs. $81\pm19$ expected 
(class 2), and 90 observed vs. $87\pm18$ expected (class 3). 
The 2-D distribution of dijet mass vs. $Q\equiv m(\rho_T)-m(\pi_T)-m(W)$ 
is used as a discriminant to set limits on the masses of techni-pion and 
techni-rho. Figure~\ref{fig:MxFVTC} shows the excluded region in 
the $m(\rho_T)-m(\pi_T)$ plane assuming the 
Technicolor Strawman model~\cite{Lane:1991qh}; the results of the 
three $b$-tagging categories are combined.

\subsection{\boldmath Search for Long-lived Particles Decaying into $ee$ or $\g\g$}
\begin{figure}
\begin{tabular}{cc}
\includegraphics[width=0.4\textwidth]{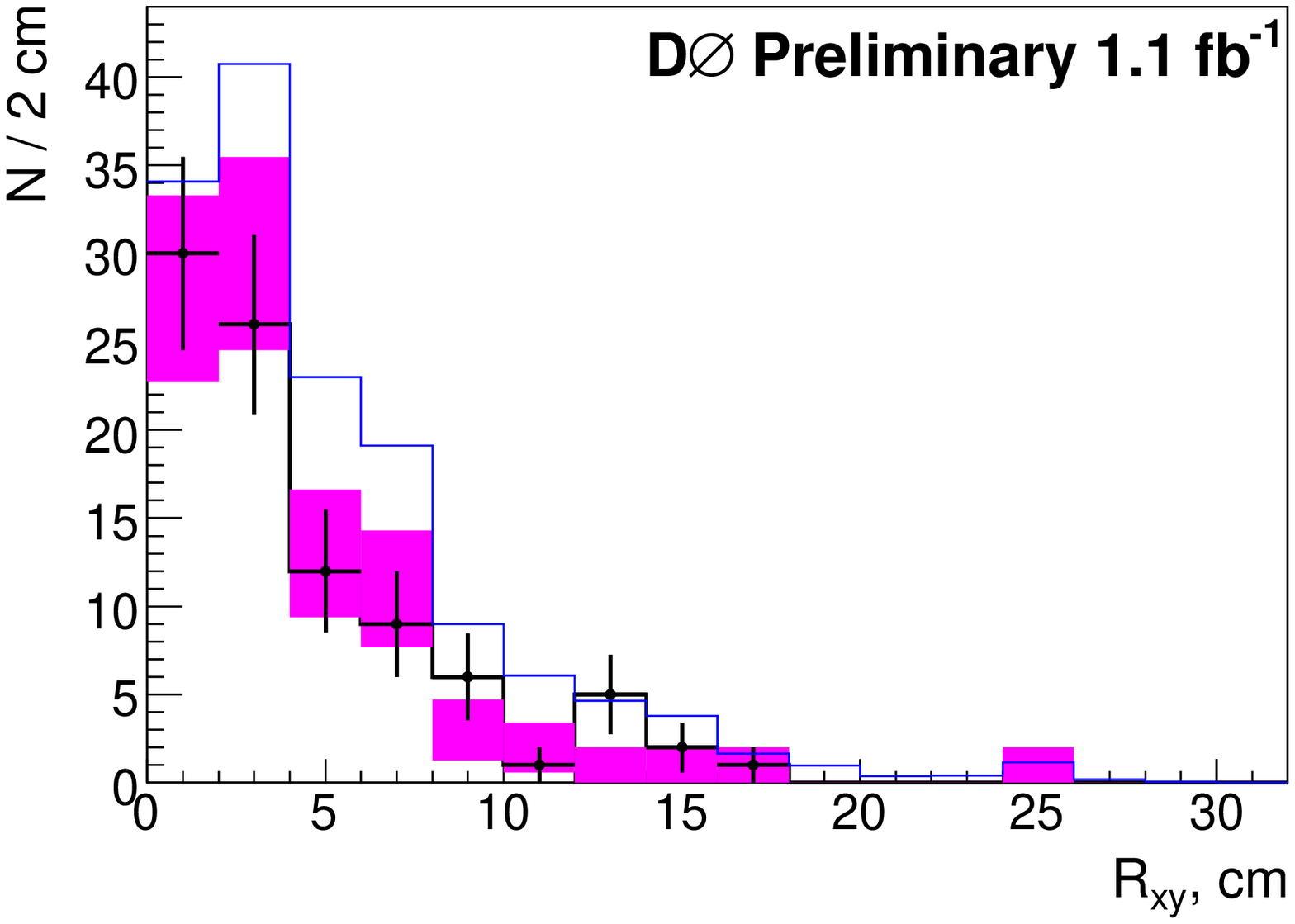} &   
\includegraphics[width=0.4\textwidth]{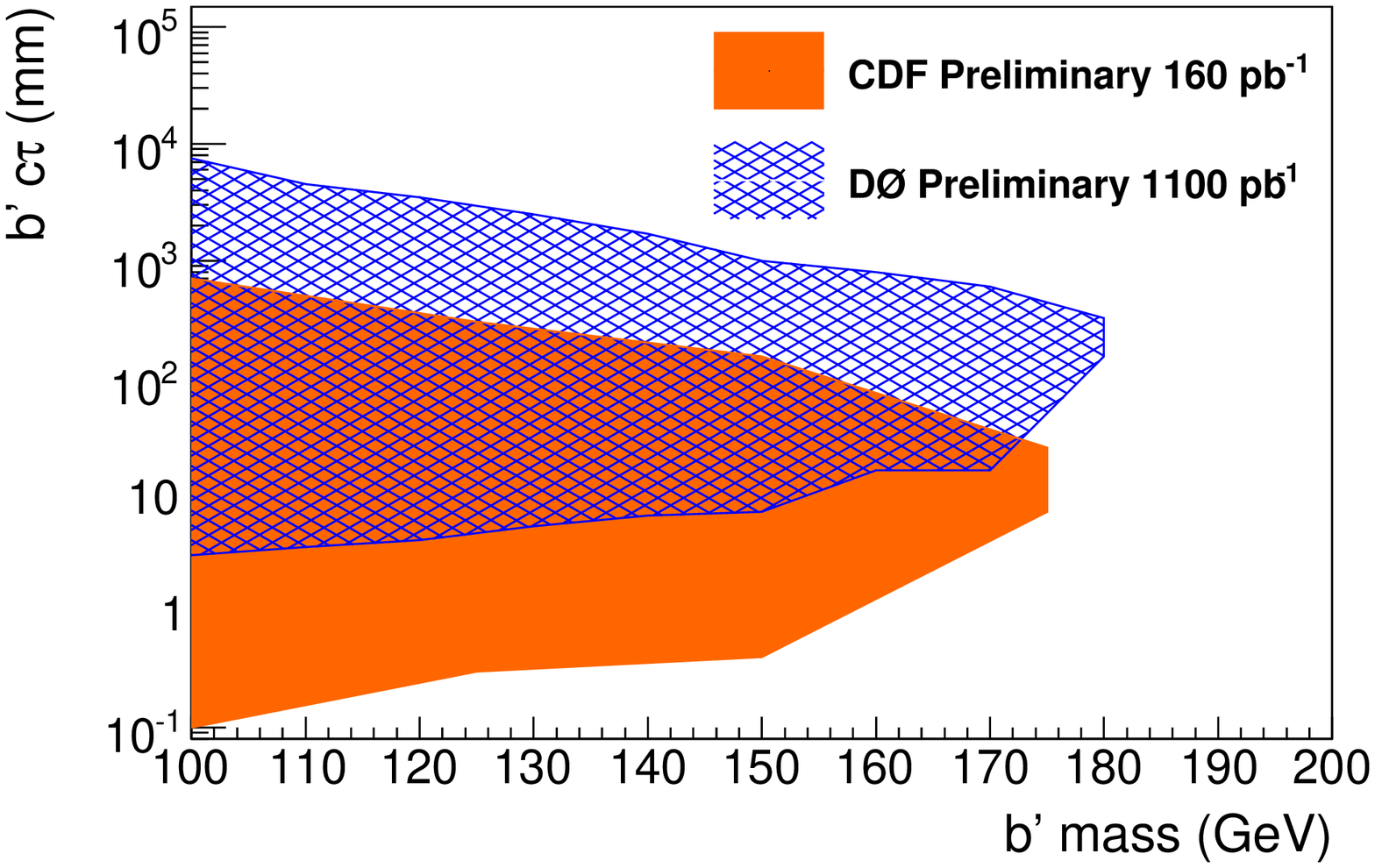} \\
\end{tabular} 
	\caption{\label{fig:LvZ}
	D0 Search for Long-lived Particles Decaying into $ee$ or $\g\g$.
	The left figure shows the $R_{xy}$ distribution observed (markers),
  	and background prediction with error bars (filled histogram). 
	The background prediction is a mirror image of the negative 
	$R_{xy}$ in data given that SM background is symmetric about 
	$R_{xy}=0$.
	The expected $R_{xy}$ for a $b^{\prime}$ with mass at 160~\gevcsq\ 
	and $c\tau$ at 300~mm is also shown. 
	The right figure shows the excluded regions in the 
	$c\tau(b^{\prime})-m(b^{\prime})$ plane by D0 (hatched) and 
	a preliminary CDF search 
	for long-lived particles decaying to $Z\rightarrow \mu\mu$ (filled).
	}
\end{figure}

\unboldmath
The D0 collaboration has looked for long-lived particles that decay into 
final states with two electrons or two photons in 1.1~\fbarn\ of data, i.e. 
a pair of EM showers that originate from the same point in space, away from 
the $p\bar{p}$ interaction point~\cite{D0LvZ}. Such long-lived particles arise 
in fourth generation ($b^{\prime}$)~\cite{bprime}, gauge-mediated SUSY breaking~\cite{LvGMSB}, and hidden valleys~\cite{HiddenValley}. 
Events selected have two central EM clusters with $\et > 20~\gev$. 
This analysis uses the ``EM pointing algorithm'' as described in 
Section~\ref{sec:LED} to find the intersection of the trajectories of these 
two EM objects (secondary vertex). An excess in the positive $R_{xy}$ compared to the negative $R_{xy}$ indicates the existence of long-lived exotic 
particles, where $R_{xy}$ is the transverse radius from the detector 
center to the secondary vertex (see Figure~\ref{fig:LvZ}). No excess is 
observed and Figure~\ref{fig:LvZ} shows the 95\% CL limits on the $c\tau$ 
and mass of the fourth generation quark $b^{\prime}$. This D0 
search is particular sensitive to $b^{\prime}$ with large life time 
($c\tau \sim 5~\mathrm{mm}-5000~\mathrm{mm}$) while 
a previous CDF search using $\mu\mu$ final state~\cite{CDFLvZ} is sensitive 
to $b^{\prime}$ with small life time 
($c\tau \sim 0.5~\mathrm{mm}-500~\mathrm{mm}$). The two analyses are 
complementary to each other.

\section{CONCLUSION}
The CDF and D0 collaborations have performed extensive signature-based searches
 and searches inspired by non-SUSY models. 
We have not yet found significant excess in 1.0--2.5~\fbarn\ of data. 
However, the result of the CDF search for high-mass $ee$ resonances 
is exciting: a 3.8~$\sigma$ excess is observed in the region 
$228 < M_{ee} < 250~\gevcsq$ with a $p$-value of 0.6\%. 
The same analysis will be updated with more data. In addition, similar 
searches in the $\mu\mu$ channel by both CDF and D0 are 
expected in the near future and will help understanding whether the 
excess is an indication of new physics or a statistical fluctuation. 
Moreover, several novel detectors and techniques have been developed, 
such as the CDF EM timing system, $\met\;$ significance model, and the D0 EM 
pointing algorithm. These allow us to explore signatures which were considered 
difficult before. As more data data are being 
collected, we expect many new and interesting results from both CDF and D0.

\begin{acknowledgments}
The author wishes to thank the CDF and D0 exotic group conveners, 
B. Brau, C. Hays, T. Adams, and P. Verdier, for their suggestions of 
the presentation in the conference. The author also would like to 
thank the following people for answering author's almost non-stop questions: 
R.~Culbertson, H.J.~Frisch, D.~Krop, C.~Pilcher, 
S.~Wilbur, A.~Loginov, I.~Shreyber, J.~Adelman, A.~Pronko, M.~Goncharov, 
C.~Henderson, G.~Choudalakis, V. Krutelyov,
Y.~Gershtein, Y. Maravin, B.R.~Ko, A.~Das, K.~Hatakeyama, C.~Magass, 
P.H. Beauchemin, E.~James, Y. Hu, Y. Nagai, W.-M. Yao, and S. Bar-Shalom.
This work has been supported by U.S. Department of Energy.

\end{acknowledgments}


\begin{thebibliography}{99}
\bibitem{sigma_ttg} 
The leading-order (LO) production cross-section of $t\bar{t}\g$ has been 
estimated using the {\sc Madgraph} generator to be 0.076~pb. A $k$-factor to 
scale the LO to the NLO cross-section, $1.10\pm 0.015$, is used after a 
discussion with F.~Petriello and U.~Baur.
\end{thebibliography}
\end{document}